\documentclass[12pt,a4paper]{article}
\pdfoutput=1
\usepackage{epsfig}
\usepackage{amssymb}
\usepackage{graphicx}
\usepackage{color}
\usepackage{subfigure}
\usepackage{mathtools}
\usepackage[hidelinks]{hyperref}

\makeatletter
\renewcommand{\theequation}{\thesection.\arabic{equation}}
\@addtoreset{equation}{section} \makeatother
\renewcommand{\vec}[1]{\underline{#1}}

\newcommand{\sss}[1]{\scriptscriptstyle{#1}}
\newcommand{\lra}[1]{\la{#1}\ra}

\def\a{\alpha}
\def\b{\beta}

\def\ep{\epsilon}
\def\e{\eta}
\def\ph{\phi}
\def\Ph{\Phi}

\def\l{\lambda}

\def\m{\mu}

\def\th{\theta}

\def\s{\sigma}
\def\S{\Sigma}
\def\ta{\tau}
\def\o{\omega}

\def\z{\zeta}

\def\lt{\left}
\def\rt{\right}

\def\nn{\nonumber}

\DeclareMathOperator{\tr}{Tr}

\def\p{\partial}

\def\la{\langle}
\def\ra{\rangle}

\def\mca{\mathcal{A}}
\def\mcb{\mathcal{B}}
\def\mcs{\mathcal{S}}
\def\mcm{\mathcal{M}}

\def\nn{\nonumber}
\def\bul{\noindent$\bullet$}

\def\bea{\begin{eqnarray}}
\def\eea{\end{eqnarray}}

\begin{document}

\begin{titlepage}
\title{\vskip -60pt
\vskip 20pt %
Junctions of mass-deformed nonlinear sigma models on $SO(2N)/U(N)$ and $Sp(N)/U(N)$ II
}
\author{
Taegyu Kim$^a$\footnote{e-mail : taegyukim@skku.edu}~ and 
Sunyoung Shin$^b$\footnote{e-mail : sihnsy@gmail.com}}
\date{}
\maketitle \vspace{-1.0cm}
\begin{center}
~~~
$^a$\it Department of Physics, Sungkyunkwan University, Suwon 16419, Republic of Korea \\
$^b$\it Department of Physics, Kangwon National University, Chuncheon 24341,\\ Republic of Korea

\end{center}

\thispagestyle{empty}

\begin{abstract}
 We study vacua, walls and three-pronged junctions of mass-deformed nonlinear sigma models on $SO(2N)/U(N)$ and $Sp(N)/U(N)$ for generic $N$. We review and discuss the on-shell component Lagrangians of the ${\mathcal{N}}=2$ nonlinear sigma model on the Grassmann manifold, which are obtained in the ${\mathcal{N}}=1$ superspace formalism and in the harmonic superspace formalism. We also show that the K\"{a}hler potential of the ${\mathcal{N}}=2$ nonlinear sigma model on the complex projective space, which is obtained in the projective superspace formalism, is equivalent to the K\"{a}hler potential of the ${\mathcal{N}}=2$ nonlinear sigma model with the Fayet-Iliopoulos parameters $c^a=(0,0,c=1)$ on the complex projective space, which is obtained in the ${\mathcal{N}}=1$ superspace formalism.
\end{abstract} 

\end{titlepage}

\vfill
\newpage
\setcounter{page}{1}
\setcounter{footnote}{0}
\renewcommand{\thefootnote}{\arabic{footnote}}

\section{Introduction} \label{sec:intro}
\setcounter{equation}{0}
%
%
Topological solitons of supersymmetric theories are Bogomol'nyi-Prasad-Sommerfield (BPS) states, which preserve a fraction of supersymmetry \cite{Witten:1978mh}. Walls and junctions preserve 1/2 SUSY \cite{Abraham:1992vb} and 1/4 SUSY respectively \cite{Abraham:1990nz}. Potential terms and various wall configurations are discussed in \cite{AlvarezGaume:1983ab,Gauntlett:2000bd,Gauntlett:2000ib,Arai:2002xa,Eto:2020vjm}. 
 
The moduli matrix formalism is proposed in  \cite{Isozumi:2004jc} to analyse walls in the ${\mathcal{N}}=2$ supersymmetric $U(N_C)$ gauge theory. In the strong coupling limit, the model becomes the mass-deformed ${\mathcal{N}}=2$ nonlinear sigma model on the Grassmann manifold $G_{N_F,N_C}$, which is defined by $G_{N+M,M}=\frac{SU(N+M)}{SU(N)\times SU(M) \times U(1)}$. The ${\mathcal{N}}=2$ nonlinear sigma model on the Grassmann manifold with the Fayet-Iliopoulos (FI) parameters $c^a=(0,0,c)$ has a bundle structure so that the model can be parametrised by a field that parametrises the base manifold and a field that parametrises the cotangent space as the fiber \cite{Arai:2002xa,Lindstrom:1983rt}. It is shown in \cite{Isozumi:2004jc} that with the FI parameters, only one of the fields in the hypermultiplet contributes to the configuration of vacua and walls. It is also shown that wall solutions are exact in the strong coupling limit. Hermitian symmetric spaces $SO(2N)/U(N)$ and $Sp(N)/U(N)$ are quadrics of the Grassmann manifold \cite{Higashijima:1999ki}. Kink solutions are studied in mass-deformed nonlinear sigma models on $SO(2N)/U(N)$ and $Sp(N)/U(N)$ \cite{Arai:2011gg,Eto:2011cv,Lee:2017kaj,Arai:2018tkf}.

Complex mass parameters can be introduced to hypermultiplets of ${\mathcal{N}}=2$ nonlinear sigma models in dimensions $D\leq 3+1$. We can construct wall junctions, which preserve 1/4 supersymmetry in mass-deformed ${\mathcal{N}}=2$ nonlinear sigma models. As only one complex field of the hypermultiplet contributes to vacua, walls and junctions of the mass-deformed ${\mathcal{N}=2}$ nonlinear sigma model on the Grassmann manifold with the FI parameters $c^a=(0,0,c)$, mass-deformed nonlinear sigma models, which have junctions, on the Grassmann manifold and the quadrics of the Grassmann manifold can be obtained by either setting the other complex field zero in ${\mathcal{N}}=2$ nonlinear sigma models or complexifying mass matrices and adjoint scalar fields in ${\mathcal{N}}=1$ nonlinear sigma models \cite{Eto:2005cp,Eto:2005fm}. 

Junctions of mass-deformed nonlinear sigma models on the complex projective space and on the Grassmann manifold are studied in the moduli matrix formalism \cite{Eto:2005cp,Eto:2005fm,Shin:2018chr}. In \cite{Eto:2005fm}, three-pronged junctions of the mass-deformed nonlinear sigma model on the Grassmann manifold are constructed by embedding the Grassmann manifold to the complex projective space using the Pl\"{u}cker embedding. In \cite{Shin:2018chr}, three-pronged junctions of the mass-deformed nonlinear sigma models on the Grassmann manifold are constructed by utilising diagrams of the pictorial representation, which is proposed in \cite{Lee:2017kaj}. In \cite{Kim:2019jzo}, three-pronged junctions of the mass-deformed nonlinear sigma models on $SO(8)/U(4)$ and $Sp(3)/U(3)$ are constructed by the diagram method, which is proposed in \cite{Shin:2018chr}.

In this paper, we construct three-pronged junctions of $SO(2N)/U(N)$ and $Sp(N)/U(N)$ for generic $N$ as a sequel to \cite{Kim:2019jzo}. We review and discuss the on-shell component Lagrangians of the ${\mathcal{N}}=2$ nonlinear sigma model on the Grassmann manifold obtained in the ${\mathcal{N}}=1$ superspace formalism and in the harmonic superspace formalism. We compare the K\"{a}hler potentials of the ${\mathcal{N}}=2$ nonlinear sigma model on the complex projective space obtained in the ${\mathcal{N}}=1$ superspace formalism and in the projective superspace formalism. 

This paper is organised as follows. In Section \ref{sec:model}, we construct mass-deformed nonlinear sigma models on $SO(2N)/U(N)$ and $Sp(N)/U(N)$ by complexifying the mass matrix and the adjoint scalar field of the mass-deformed ${\mathcal{N}}=1$ supersymmetric nonlinear sigma models on $SO(2N)/U(N)$ and $Sp(N)/U(N)$ to study vacua, walls and three-pronged junctions. In Section \ref{sec:so}, we study three-pronged junctions of mass-deformed nonlinear sigma models on $SO(2N)/U(N)$ with $N=4m$ and $N=4m+1$. In Section \ref{sec:sp}, we study three-pronged junctions of mass-deformed nonlinear sigma models on $Sp(N)/U(N)$ with $N=4m-1$ and $N=4m$. In Section \ref{sec:tastgr}, we review and discuss the ${\mathcal{N}}=2$ nonlinear sigma models on the Grassmann manifold or the complex projective space. We show that the on-shell component Lagrangian of the nonlinear sigma model on the Grassmann manifold obtained in the harmonic superspace formalism \cite{Galperin:2001uw,Galperin:1985dw} is equivalent to the on-shell component Lagrangian obtained in the ${\mathcal{N}}=1$ superspace formalism \cite{Lindstrom:1983rt,Rocek:1980kc}. We also show that the K\"{a}hler potential of the ${\mathcal{N}}=2$ nonlinear sigma model on the complex projective space obtained in the projective superspace formalism \cite{Arai:2006gg} is equivalent to the K\"{a}hler potential of the ${\mathcal{N}}=2$ nonlinear sigma model with the Fayet-Iliopoulos parameters $c^a=(0,0,c=1)$ on the complex projective space obtained in the ${\mathcal{N}}=1$ superspace formalism \cite{Arai:2002xa,Lindstrom:1983rt}.

\section{Models} \label{sec:model}
\setcounter{equation}{0}

Vacua and walls of mass-deformed nonlinear sigma models on $SO(2N)/U(N)$ and $Sp(N)/U(N)$ are constructed in the ${\mathcal{N}}=1$ supersymmetric nonlinear sigma models \cite{Arai:2011gg,Lee:2017kaj,Arai:2018tkf}. The bosonic part of the Lagrangian is
\begin{align}\label{eq:n1lag}
{\mathcal{L}}=\tr\lt(D_\m\mca\overline{D^\m\mca}-|\mca {\widetilde{M}} -\S\mca|^2-|2J\mca^T\mca_0|^2\rt),
\end{align} 
with constraints
\begin{align}\label{eq:lagconst}
&\mca\bar{\mca}-cI_M=0, \nn\\
&\mca J\mca^T=0,\quad \mathrm{(c.t.)}=0,\nn\\
&J=\lt\{\begin{array}{c}
\s^1\otimes I_N,\quad SO(2N)/U(N)\\
i\s^2\otimes I_N,\quad Sp(N)/U(N).
\end{array}\rt.
\end{align}
We follow the convention of \cite{Galperin:2001uw}. $\mca$ is an $N\times 2N$ matrix field, $\widetilde{M}$ is a real-valued $2N\times 2N $ mass matrix, $\S$ is a real-valued $N\times N$ matrix field and $\mca_0$ is an $N\times N$ matrix field. We diagonalise $\S$ for later use.

We construct the bosonic component Lagrangian of mass-deformed nonlinear sigma models on $SO(2N)/U(N)$ and $Sp(N)/U(N)$ to study three-pronged junctions. The Lagrangian of the mass-deformed nonlinear sigma model, which has three-pronged junctions, on the Grassmann manifold can be obtained by replacing the mass matrix and the adjoint scalar field of the ${\mathcal{N}}=1$ model with a complex valued mass matrix and a complex-valued matrix field \cite{Eto:2005cp,Eto:2005fm}. Therefore we can obtain the mass-deformed nonlinear sigma models on $SO(2N)/U(N)$ and $Sp(N)/U(N)$, which are quadrics of the Grassmann manifold, by replacing the mass matrix $\widetilde{M}$ and the scalar field $\S$ with a complex valued mass matrix $M$ and a complex-valued matrix field $\mcs$: 
\begin{align}\label{eq:n2lag}
{\mathcal{L}}=\tr\lt(D_\m\mca\overline{D^\m\mca}-|\mca M -\mcs\mca|^2-|2J\mca^T\mca_0|^2\rt).
\end{align} 
This type of complexification is also considered to construct dyonic configurations with non-parallel charge vectors \cite{Eto:2011cv}. It is also shown that the Lagrangian can be derived from a recently proposed nonlinear sigma model \cite{Kim:2019jzo}. The Lagrangian (\ref{eq:n2lag}) is constrained by (\ref{eq:lagconst}).

The complex mass matrix $M$ is a linear combination of the Cartan generators $H_i$, $(i=1,\cdots,N)$:
\begin{align}\label{eq:cplxmass}
&M=\vec{l}\cdot\vec{H}, \nn\\
&\vec{l}:=(m_{\sss{1}}+in_{\sss{1}},\cdots,m_{\sss{N}}+in_{\sss{N}}), \nn\\
&\vec{H}:=(H_1,\cdots,H_{\sss{N}}).
\end{align}
The matrix field $\mcs$ can be parametrised by real-valued matrices $\s_i$ and $\ta_i$, $(i=1,\cdots,N)$:
\begin{align}\label{eq:cplxsfld}
\mcs=\mathrm{diag}(\s_{\sss{1}}+i\ta_{\sss{1}},\cdots,\s_{\sss{N}} +i\ta_{\sss{N}}).
\end{align}
The vacuum equations are
\begin{align}
&\mca M -\mcs \mca =0, \nn\\
&\mca^T\mca_0=0.
\end{align}
The vacua are labelled by
\begin{align}
(\s_{\sss{1}}+i\ta_{\sss{1}},\cdots,\s_{\sss{N}}+i\ta_{\sss{N}})
=(\pm(m_{\sss{1}}+in_{\sss{1}}),\cdots,\pm(m_{\sss{N}}+in_{\sss{N}})).
\end{align}
There are $2^{N-1}$ vacua in the mass-deformed nonlinear sigma model on $SO(2N)/U(N)$ and there are $2^N$ vacua in the mass-deformed nonlinear sigma model on $Sp(N)/U(N)$ \cite{Arai:2011gg,Lee:2017kaj,Arai:2018tkf}.

In this paper, we consider the following Lagrangian:
\begin{align}\label{eq:mainlag}
{\mathcal{L}}=\tr\lt(D_\m\mca\overline{D^\m\mca}-\sum_{a=1,2}|\mca M_a -\S_a\mca|^2-|2J\mca^T\mca_0|^2\rt),
\end{align} 
where $M_a$ and $\S_a$, $(a=1,2)$ are real valued, which are related to $M$ and $\mcs$ by
\begin{align}
&M_1=\mathrm{Re}(M),\quad M_2=\mathrm{Im}(M),\nn\\
&\S_1=\mathrm{Re}(\mcs), \quad \S_2=\mathrm{Im}(\mcs).
\end{align}

We investigate the static configurations, which have the Poincar\'{e} invariance on the worldvolume: $\p_0=\p_3=0$, $A_0=A_3=0$. The energy density can be calculated from the Lagrangian (\ref{eq:mainlag}). The Bogomol'nyi completion of the energy density produces the BPS equation. The solution to the BPS equation in the moduli matrix formalism is
\begin{align}\label{eq:bps_sol}
\mca=S^{-1}H_0e^{M_1x^1+M_2x^2}.
\end{align}
The matrix $H_0$ is the moduli matrix. The constraints (\ref{eq:lagconst}) correspond to the following equations:
\begin{align}\label{eq:const_moduli}
&S\bar{S}=\frac{1}{c}H_0e^{2M_1x^1+2M_2x^2}\bar{H}_0, \nn\\
&H_0 J H^{T}_0=0.
\end{align}
The worldvolume symmetry is
\begin{align}\label{eq:wvs}
H_0^\prime = V H_0,\quad S^\prime=V S, \quad V\in GL(N,\mathbf{C}).
\end{align}
The first equation in (\ref{eq:wvs}) and the second equation (\ref{eq:const_moduli}) show that the moduli matrices parametrise $SO(2N)/U(N)$ and $Sp(N)/U(N)$ respectively \cite{Arai:2011gg,Lee:2017kaj,Arai:2018tkf}.

Walls are built from elementary walls, which are identified with scaled simple roots of the global symmetry. The simple root generators and the simple roots are summarised below \cite{Eto:2011cv,Isaev:2018xcg}:\\

\bul $SO(2N)$
\begin{align}\label{eq:gen_rt_so}
&E_i=e_{i,i+1}-e_{i+{\sss{N}}+1,i+{\sss{N}}},\quad (i=1,\cdots,N-1), \nn\\
&E_N=e_{\sss{N-1,2N}}-e_{\sss{N,2N-1}}, \nn\\
&\vec{a}_i=\hat{e}_i-\hat{e}_{i+1},  \nn\\
&\vec{a}_{\sss{N}}=\hat{e}_{\sss{N-1}}+\hat{e}_{\sss{N}}.
\end{align}

\bul $USp(2N)$
\begin{align}\label{eq:gen_rt_sp}
&E_i=e_{i,i+1}-e_{i+{\sss{N}}+1,i+{\sss{N}}},\quad (i=1,\cdots,N-1), \nn\\
&E_N=e_{\sss{N,2N}}, \nn\\
&\vec{a}_i=\hat{e}_i-\hat{e}_{i+1},  \nn\\
&\vec{a}_{\sss{N}}=2\hat{e}_{\sss{N}}.
\end{align}

\section{Three-pronged junctions of the mass-deformed nonlinear sigma models on $SO(2N)/U(N)$} \label{sec:so}
\setcounter{equation}{0}

\begin{figure}[ht!]
\vspace{1cm}
\begin{center}
$\begin{array}{ccc}
\includegraphics[width=1.5cm,clip]{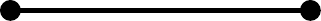}
&~~~~~~&
\includegraphics[width=4cm,clip]{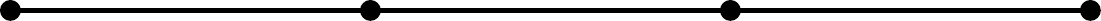}\\
\mathrm{(a)} &~~~~~~& \mathrm{(b)} \\
~~ &  ~~ & ~~ \\
\includegraphics[width=7cm,clip]{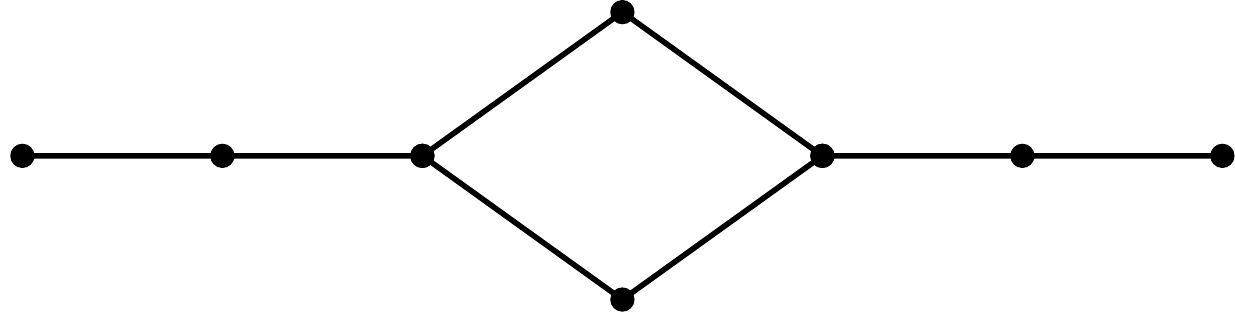}
&~~~~~~&
\includegraphics[width=7cm,clip]{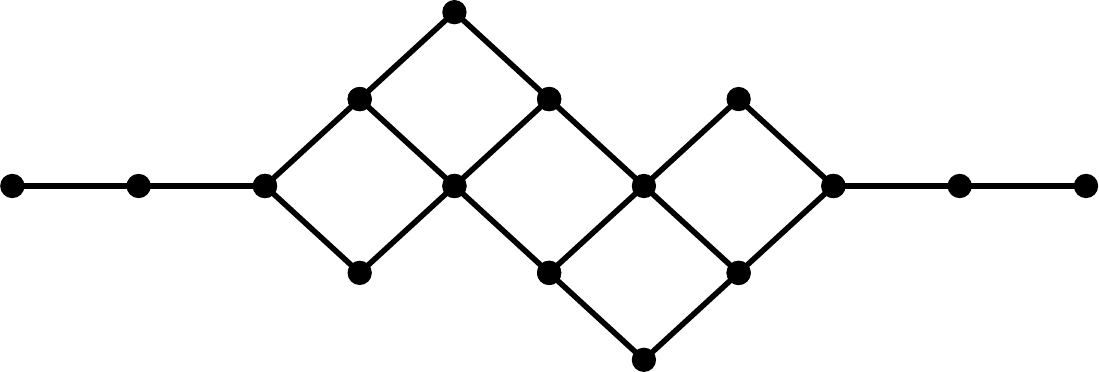}\\
\mathrm{(c)} &~~~~~~& \mathrm{(d)}
\end{array}
$
\end{center}
 \caption{There is a recurrence of one of the  diagrams for each $N$ mod 4. (a)$N=4m-2$, (b)$N=4m-1$, (c)$N=4m$, and (d)$N=4m+1$, $(m\geq 1)$. }
 \label{fig:son}
\end{figure}

Vacua and elementary walls of the mass-deformed nonlinear sigma models on $SO(2N)/U(N)$ are studied in the pictorial representation \cite{Lee:2017kaj}. Vacua and elementary walls are presented as vertices and segments of diagrams in the pictorial representation. All the diagrams of vacua and elementary walls are symmetric, although the array of the simple roots depends on the parity of the flavour number. There is a recurrence of a diagram for each $N$ mod 4 in the vacuum structures that are connected to the maximum number of elementary walls. The diagrams are presented in Figure \ref{fig:son}. The vacuum structures are proved by induction. The whole structure of vacua and elementary walls of the mass-deformed nonlinear sigma models on $SO(2N)/U(N)$ for generic $N$ can be derived from the vacuum structures that are connected to the maximum number of elementary walls.

In this section we construct three-pronged junctions from the diagrams (c) and (d) of Figure \ref{fig:son} for generic $N$.

The vacua are labelled in the descending order \cite{Lee:2017kaj}:
\begin{align} \label{eq:so_vac_rule}
&(\s_{\sss{1}}+i\ta_{\sss{1}},\s_{\sss{2}}+\ta_{\sss{2}},\cdots,\s_{{\sss{N-1}}}+i\ta_{{\sss{N-1}}},\s_{{\sss{N}}}+i\ta_{{\sss{N}}})\nn\\
&\quad\quad =(m_{\sss{1}}+in_{\sss{1}},m_{\sss{2}}+in_{\sss{2}},\cdots,m_{{\sss{N-1}}}+in_{{\sss{N-1}}},m_{{\sss{N}}}+in_{{\sss{N}}}),  \nn\\
&(\s_{\sss{1}}+i\ta_{\sss{1}},\s_{\sss{2}}+\ta_{\sss{2}},\cdots,\s_{{\sss{N-1}}}+i\ta_{{\sss{N-1}}},\s_{{\sss{N}}}+i\ta_{{\sss{N}}})\nn\\
&\quad\quad =(m_{\sss{1}}+in_{\sss{1}},m_{\sss{2}}+in_{\sss{2}},\cdots,-m_{{\sss{N-1}}}-in_{{\sss{N-1}}},-m_{{\sss{N}}}-in_{{\sss{N}}}),  \nn\\
&\quad \vdots \nn\\
&(\s_{\sss{1}}+i\ta_{\sss{1}},\s_{\sss{2}}+\ta_{\sss{2}},\cdots,\s_{{\sss{N-1}}}+i\ta_{{\sss{N-1}}},\s_{{\sss{N}}}+i\ta_{{\sss{N}}})\nn\\
&\quad\quad =(\pm m_{\sss{1}}\pm in_{\sss{1}},-m_{\sss{2}}-in_{\sss{2}},\cdots,-m_{{\sss{N-1}}}-in_{{\sss{N-1}}},-m_{{\sss{N}}}-in_{{\sss{N}}}),  \nn\\
\end{align}
where the sign $\pm$ is $+$ for odd $N$ and $-$ for even $N$. Let $\lra{\cdot}$ denote a vacuum, and $\lra{\cdot \leftarrow \cdot}$ or $\lra{\cdot \leftrightarrow \cdot}$ denote a wall.

\subsection{$N=4m$} \label{sec:so_n=4m}

\begin{figure}[ht!]
\begin{center}
\includegraphics[width=7cm,clip]{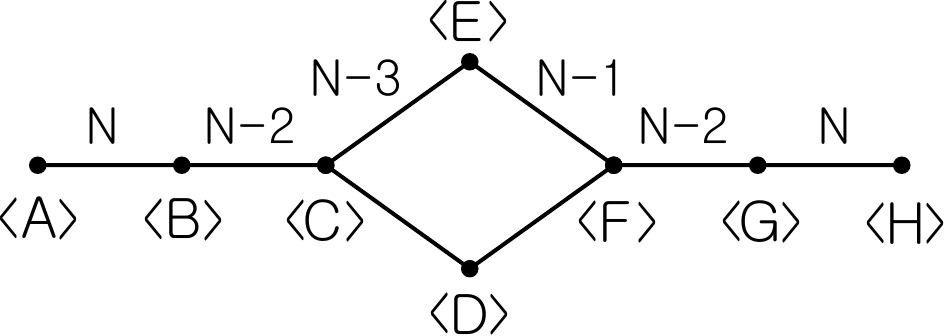}
\end{center}
 \caption{ Vacua and elementary walls of the mass-deformed nonlinear sigma model on $SO(2N)/U(N)$, $N=4m$ \cite{Lee:2017kaj}. The numbers in brackets indicate the vacuum labels. The numbers without brackets indicate the subscript $i$'s of simple roots $\vec{\a}_i$, $(i=N-3,\cdots,N)$.}
 \label{fig:son_wall_4m}
\end{figure}

In this subsection, we construct three-pronged junctions of the mass-deformed nonlinear sigma model on $SO(2N)/U(N)$ with $N=4m$, $(m\geq 1)$. We reduce the model (\ref{eq:mainlag}) by setting $M_2=0$ and $\S_2=0$ to the model (\ref{eq:n1lag}) to investigate the structure of vacua and elementary walls. The diagram in Figure \ref{fig:son_wall_4m} recurs at the vacua that are connected to the maximum number of elementary walls \cite{Lee:2017kaj}. 

Elementary walls are identified with scaled simple roots $\vec{g}_{\lra{\cdot\leftarrow\cdot}}=2c\vec{\a}_{\sss{I}}$, $(I=1,\cdots,N)$ with $\a_{\sss{I}}$ in (\ref{eq:gen_rt_so}), and compressed walls are linear combinations of the scaled simple roots \cite{Lee:2017kaj,Kim:2019jzo}. The single walls of the recurring diagram are derived below:
\begin{align} \label{eq:walls_so_a}
&\lra{A\leftarrow B}=\lra{G\leftarrow H}= 2c\vec{\a}_{\sss{N}}=2c\lt(\hat{e}_{\sss{N-1}}+ \hat{e}_{\sss{N}}\rt), \nn\\
&\lra{A\leftarrow C}=\lra{F\leftarrow H}= 2c\vec{\a}_{\sss{N-2}}+2c\vec{\a}_{\sss{N}}=2c\lt(\hat{e}_{\sss{N-2}}+\hat{e}_{\sss{N}} \rt),\nn\\
&\lra{A\leftarrow D}=\lra{E\leftarrow H}= 2c\vec{\a}_{\sss{N-2}}+2c\vec{\a}_{\sss{N-1}}+2c\vec{\a}_{\sss{N}}=2c\lt(\hat{e}_{\sss{N-2}}+ \hat{e}_{\sss{N-1}}\rt),\nn\\
&\lra{A\leftarrow E}=\lra{D\leftarrow H}= 2c\vec{\a}_{\sss{N-3}}+2c\vec{\a}_{\sss{N-2}}+2c\vec{\a}_{\sss{N}}=2c\lt(\hat{e}_{\sss{N-3}}+ \hat{e}_{\sss{N}}\rt),\nn\\
&\lra{A\leftarrow F}=\lra{C\leftarrow H}= 2c\vec{\a}_{\sss{N-3}}+2c\vec{\a}_{\sss{N-2}}+2c\vec{\a}_{\sss{N-1}}+2c\vec{\a}_{\sss{N}}=2c\lt(\hat{e}_{\sss{N-3}}+ \hat{e}_{\sss{N-1}}\rt),\nn\\
&\lra{A\leftarrow G}=\lra{B\leftarrow H}= 2c\vec{\a}_{\sss{N-3}}+4c\vec{\a}_{\sss{N-2}}+2c\vec{\a}_{\sss{N-1}}+2c\vec{\a}_{\sss{N}}=2c\lt(\hat{e}_{\sss{N-3}}+ \hat{e}_{\sss{N-2}}\rt),\nn\\
&\lra{A\leftarrow H}= 0. 
\end{align}
\begin{align}
&\lra{B\leftarrow C}=\lra{F\leftarrow G}= 2c\vec{\a}_{\sss{N-2}} =2c\lt(\hat{e}_{\sss{N-2}}-\hat{e}_{\sss{N-1}} \rt),\nn\\ 
&\lra{B\leftarrow D}= \lra{E\leftarrow G}= 2c\vec{\a}_{\sss{N-2}}+2c\vec{\a}_{\sss{N-1}}=2c \lt(\hat{e}_{\sss{N-2}}- \hat{e}_{\sss{N}} \rt),\nn\\
&\lra{B\leftarrow E}= \lra{D\leftarrow G}= 2c\vec{\a}_{\sss{N-3}}+2c\vec{\a}_{\sss{N-2}}=2c\lt(\hat{e}_{\sss{N-3}}- \hat{e}_{\sss{N-1}}\rt),\nn\\
&\lra{B\leftarrow F}= \lra{C\leftarrow G}= 2c\vec{\a}_{\sss{N-3}}+2c\vec{\a}_{\sss{N-2}}+2c\vec{\a}_{\sss{N-1}}=2c\lt(\hat{e}_{\sss{N-3}}-\hat{e}_{\sss{N}} \rt),\nn\\
&\lra{B\leftarrow G}= 0. 
\end{align}
\begin{align}
&\lra{C\leftarrow D}=\lra{E\leftarrow F}= 2c\vec{\a}_{\sss{N-1}}=2c\lt(\hat{e}_{\sss{N-1}}-\hat{e}_{\sss{N}}\rt),\nn\\
&\lra{C\leftarrow E}=\lra{D\leftarrow F}= 2c\vec{\a}_{\sss{N-3}}=2c\lt(\hat{e}_{\sss{N-3}}-\hat{e}_{\sss{N-2}}\rt),\nn\\
&\lra{C\leftarrow F}= 0, \nn\\
&\lra{D\leftarrow E}= 0.
\end{align}

The moduli matrices for the walls $\lra{A\leftarrow B}$, $\lra{B\leftarrow C}$ and $\lra{A\leftarrow C}$ can be read from the diagram in Figure \ref{fig:son_wall_4m}:
\begin{align}\label{eq:single_walls_so_a}
&H_{0\lra{A\leftarrow B}}=H_{0\lra{A}} e^{e^rE_{\sss{N}}}, \nn\\
&H_{0\lra{B\leftarrow C}}=H_{0\lra{B}} e^{e^rE_{\sss{N-2}}}, \nn\\
&H_{0\lra{A\leftarrow C}}=H_{0\lra{A}} e^{e^r[E_{\sss{N}},E_{\sss{N-2}}]}.
\end{align}
The wall operators are defined by the simple root generators in (\ref{eq:gen_rt_so}):
\begin{align} \label{eq:eee1}
&E_{\sss{N-2}}=e_{\sss{N-2,N-1}}-e_{\sss{2N-1,2N-2}}, \nn\\
&E_{\sss{N}}=e_{\sss{N-1,2N}}-e_{\sss{N,2N-1}}, \nn\\
&\lt[ E_{\sss{N}},E_{\sss{N-2}}\rt]=-e_{\sss{N-2,2N}}+e_{\sss{N,2N-2}}.
\end{align}

\begin{figure}[ht!]
\begin{center}
\includegraphics[width=9cm,clip]{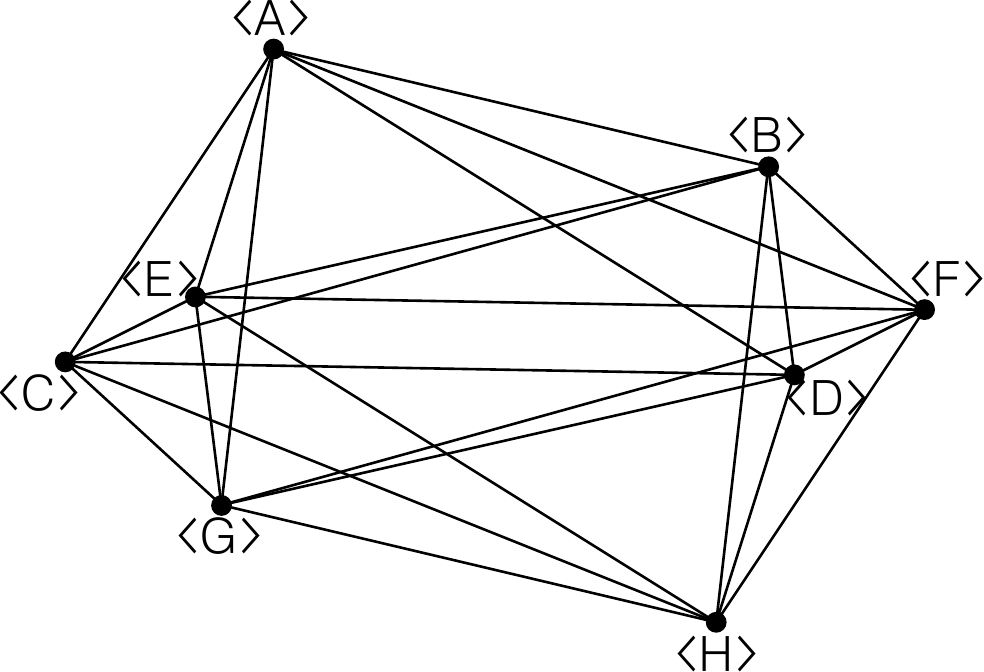}
\end{center}
 \caption{$SO(2N)/U(N)$, $N=4m$. Vertices, segments and triangular faces correspond to vacua, walls and three-pronged junctions.}
 \label{fig:son_junction_4m}
\end{figure}

We build three-pronged junctions. We set $M_2\neq 0$ and $\S_2\neq 0$ and reformulate the diagram in Figure \ref{fig:son_wall_4m} to produce the diagram in Figure \ref{fig:son_junction_4m}. Vertices, segments and triangular faces correspond to vacua, walls and three-pronged junctions. 

We study the three-pronged junction that divides the vacua $\{\lra{A},\lra{B},\lra{C}\}$. The relevant components of vacua can be read from the diagram in Figure \ref{fig:son_wall_4m}, the moduli matrices in (\ref{eq:single_walls_so_a}) and the wall operators in (\ref{eq:eee1}). A vacuum moduli matrix has a nonzero element at either $(n,n)$ or $(n,N+n)$ for each $n$-th row, $(n\leq N)$. $E_{\sss{N}}$ has nonzero $(N-1,2N)$ and $(N,2N-1)$. Therefore the vacuum $\lra{A}$ should have nonzero components in $(N-1,N-1)$ and $(N,N)$. The operator $e^{e^rE_{\sss{N}}}$ adds $(N-1,2N)$ and $(N,2N-1)$ with the opposite signs to $\lra{A}$. We can apply the same method to the walls that interpolate vacua $\lt\{\lra{B},\lra{C}\rt\}$ and $\lt\{\lra{C},\lra{A}\rt\}$. We transform the moduli matrices in (\ref{eq:single_walls_so_a}) by the worldvolume symmetry transformation (\ref{eq:wvs}) to produce generic wall solutions. The relevant elements of the moduli matrices for $\lra{A\leftrightarrow B}$, $\lra{B\leftrightarrow C}$ and $\lra{C\leftrightarrow A}$ are listed below:
\begin{align}
&H_{0\lra{A \leftrightarrow B}}^{({\sss{N-1}},{\sss{N-1}})}=H_{0\lra{A \leftrightarrow B}}^{({\sss{N}},{\sss{N}})}=\exp\lt(a_{{\sss{N-1};\sss{N-1}}}+ib_{{\sss{N-1};\sss{N-1}}}\rt),\nn\\
&H_{0\lra{A \leftrightarrow B}}^{({\sss{N-1}},{\sss{2N}})}=-H_{0\lra{A \leftrightarrow B}}^{({\sss{N}},{\sss{2N-1}})}=\exp\lt(a_{{\sss{N-1};\sss{2N}}}+ib_{{\sss{N-1};\sss{2N}}}\rt),\label{eq:ab_so}\\
& ~ \nn\\
&H_{0\lra{B \leftrightarrow C}}^{({\sss{N-2}},{\sss{N-2}})}=H_{0\lra{B \leftrightarrow C}}^{({\sss{N-1}},{\sss{2N-1}})}=\exp\lt(a_{{\sss{N-2};\sss{N-2}}}+ib_{{\sss{N-2};\sss{N-2}}}\rt),\nn\\
&H_{0\lra{B \leftrightarrow C}}^{({\sss{N-2}},{\sss{N-1}})}=-H_{0\lra{B \leftrightarrow C}}^{({\sss{N-1}},{\sss{2N-2}})}=\exp\lt(a_{{\sss{N-2};\sss{N-1}}}+ib_{{\sss{N-2};\sss{N-1}}}\rt), \label{eq:bc_so}\\
& ~ \nn\\
&H_{0\lra{C \leftrightarrow A}}^{({\sss{N-2}},{\sss{N-2}})}=H_{0\lra{C \leftrightarrow A}}^{({\sss{N}},{\sss{N}})}=\exp\lt(a_{{\sss{N-2};\sss{N-2}}}+ib_{{\sss{N-2};\sss{N-2}}}\rt),\nn\\
&H_{0\lra{C \leftrightarrow A}}^{({\sss{N-2}},{\sss{2N}})}=-H_{0\lra{C \leftrightarrow A}}^{({\sss{N}},{\sss{2N-2}})}=\exp\lt(a_{{\sss{N-2};\sss{2N}}}+ib_{{\sss{N-2};\sss{2N}}}\rt). \label{eq:ca_so}
\end{align}
The first superscript is the row number and the second superscript is the column number.

The wall solutions are obtained by the relation (\ref{eq:bps_sol}). The relevant elements of the walls are presented in \ref{app:so8m}.

The walls are located at the positions where the real parts of the elements with the same colour number are equal. The position of $\lra{A\leftrightarrow B}$ is
\begin{align}
&\mathrm{Re}\lt(\mca_{\lra{A \leftrightarrow B}}^{({\sss{N-1}},{\sss{N-1}})}\rt)
=\mathrm{Re}\lt(\mca_{\lra{A \leftrightarrow B}}^{({\sss{N-1}},{\sss{2N}})}\rt), \nn\\
&\mathrm{Re}\lt(\mca_{\lra{A \leftrightarrow B}}^{({\sss{N}},{\sss{N}})}\rt)
=\mathrm{Re}\lt(\mca_{\lra{A \leftrightarrow B}}^{({\sss{N}},{\sss{2N-1}})}\rt),\\
& \lt(m_{\sss{N-1}}+ m_{\sss{N}}\rt)x^1+\lt(n_{\sss{N-1}}+ n_{\sss{N}}\rt)x^2 +
\lt(a_{\sss{N-1;N-1}}-a_{\sss{N-1;2N}}\rt)=0.\label{eq:pos_ab_so}
\end{align}
Only one of the parameters $a_{{\sss{N-1;I}}}$, $(I=N-1,2N)$ is independent since the moduli matrices in (\ref{eq:ab_so}) have the worldvolume symmetry (\ref{eq:wvs}).

The position of $\lra{B\leftrightarrow C}$ is
\begin{align}
&\mathrm{Re}\lt(\mca_{\lra{B \leftrightarrow C}}^{({\sss{N-2}},{\sss{N-2}})}\rt)
=\mathrm{Re}\lt(\mca_{\lra{B \leftrightarrow C}}^{({\sss{N-2}},{\sss{N-1}})}\rt),\nn\\
&\mathrm{Re}\lt(\mca_{\lra{B \leftrightarrow C}}^{({\sss{N-1}},{\sss{2N-2}})}\rt) 
=\mathrm{Re}\lt(\mca_{\lra{B \leftrightarrow C}}^{({\sss{N-1}},{\sss{2N-1}})}\rt),\\
&\lt(m_{\sss{N-2}}-m_{\sss{N-1}}\rt)x^1 +\lt(n_{\sss{N-2}}-n_{\sss{N-1}}\rt)x^2 +\lt(a_{\sss{N-2;N-2}}-a_{\sss{N-2;N-1}}\rt)=0. \label{eq:pos_bc_so}
\end{align}
Only one of the parameters $a_{{\sss{N-2;I}}}$, $(I=N-2,N-1)$ is independent.

The position of $\lra{C \leftrightarrow A}$ is
\begin{align}
&\mathrm{Re}\lt(\mca_{\lra{C \leftrightarrow A}}^{({\sss{N-2}},{\sss{N-2}})}\rt)
=\mathrm{Re}\lt(\mca_{\lra{C \leftrightarrow A}}^{({\sss{N-2}},{\sss{2N}})}\rt),\nn\\
&\mathrm{Re}\lt(\mca_{\lra{C \leftrightarrow A}}^{({\sss{N}},{\sss{N}})}\rt)
=\mathrm{Re}\lt(\mca_{\lra{C \leftrightarrow A}}^{({\sss{N}},{\sss{2N-2}})}\rt),\\
&\lt( m_{\sss{N-2}}+m_{\sss{N}} \rt)x^1 + \lt( n_{\sss{N-2}}+n_{\sss{N}} \rt)x^2
+\lt(a_{\sss{N-2;N-2}}-a_{\sss{N-2;2N}}\rt)=0. \label{eq:pos_ca_so}
\end{align}
Only one of the parameters $a_{{\sss{N-2;I}}}$, $(I=N-2,2N)$ is independent.

There is a condition:
\begin{align} \label{eq:so_abc_cond}
(a_{\sss{N-2;N-1}}-a_{\sss{N-2;2N}})-(a_{\sss{N-1;N-1}}-a_{\sss{N-1;2N}})=0.
\end{align}
Therefore there are only two independent parameters to specify the junction position as expected. The position of the junction that divides $\lt\{\lra{A},\lra{B},\lra{C}\rt\}$ is calculated from (\ref{eq:pos_ab_so}), (\ref{eq:pos_bc_so}) and (\ref{eq:pos_ca_so}):
\begin{align}\label{eq:so_abc_pos}
&(x,y)=\lt(\frac{S_1}{S_3},\frac{S_2}{S_3}\rt),\nn\\
&S_1=(a_{\sss{N-2;N-2}}-a_{\sss{N-2;N-1}})(n_{\sss{N-1}}+n_{\sss{N}}) \nn\\
&\quad\quad  -(a_{\sss{N-1;N-1}}-a_{\sss{N-1;2N}})(n_{\sss{N-2}}-n_{\sss{N-1}}), \nn\\
&S_2=-(a_{\sss{N-2;N-2}}-a_{\sss{N-2;N-1}})(m_{\sss{N-1}}+m_{\sss{N}}) \nn\\
&\quad\quad+(a_{\sss{N-1;N-1}}-a_{\sss{N-1;2N}})(m_{\sss{N-2}}-m_{\sss{N-1}}),\nn\\
&S_3=-m_{\sss{N-2}}(n_{\sss{N-1}}+n_{\sss{N}})+m_{\sss{N-1}}(n_{\sss{N-2}}+n_{\sss{N}})
+m_{\sss{N}}(n_{\sss{N-2}}-n_{\sss{N-1}}).
\end{align}
The solution (\ref{eq:so_abc_pos}) is constrained by (\ref{eq:so_abc_cond}).

\subsection{$N=4m+1$} \label{sec:so_n=4m+1}

\begin{figure}[ht!]
\begin{center}
\includegraphics[width=10cm,clip]{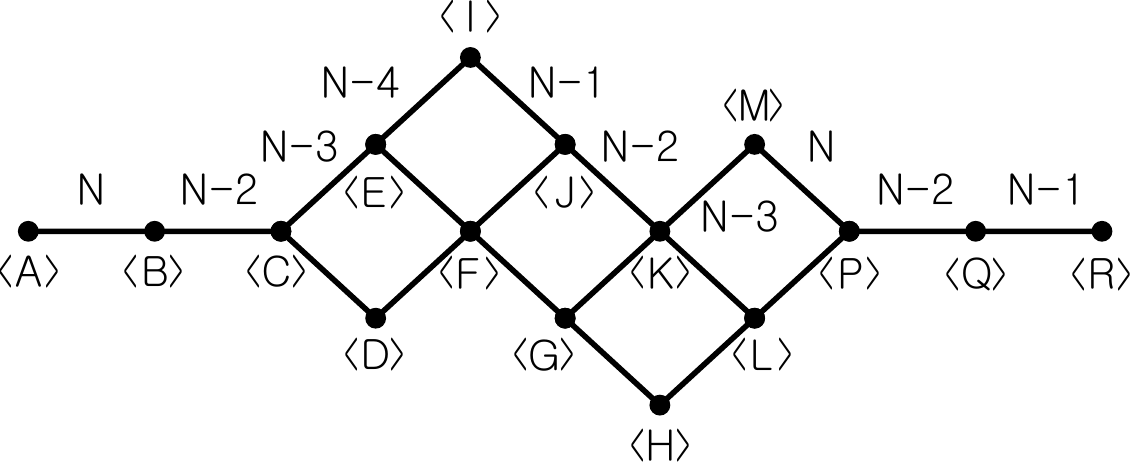}
\end{center}
 \caption{ Vacua and elementary walls of the mass-deformed nonlinear sigma model on $SO(2N)/U(N)$, $N=4m+1$ \cite{Lee:2017kaj}. The numbers in brackets indicate the vacuum labels. The numbers without brackets indicate the subscript $i$'s of simple roots $\vec{\a}_i$, $(i=N-4,\cdots,N)$.}
 \label{fig:son_wall_4m+1}
\end{figure}

In this subsection, we construct three-pronged junctions of the mass-deformed nonlinear sigma model on $SO(2N)/U(N)$ with $N=4m+1$. The structure of vacua and walls can be analysed in the reduced mass-deformed model with $M_2=0$ and $\S_2=0$. There is a recurrence of the diagram in Figure \ref{fig:son_wall_4m+1}  at the vacua that are connected to the maximum number of elementary walls \cite{Lee:2017kaj}.
The single walls that interpolate the vacua $\{\lra{H}$,$\lra{K}$,$\lra{L}$,$\lra{M}$,$\lra{P}$,$\lra{Q}$,$\lra{R}\}$ can be derived from the diagram in Figure \ref{fig:son_wall_4m+1}:
\begin{align}
&\lra{H \leftarrow K}=\lra{H \leftarrow M}=0,\nn\\
&\lra{H \leftarrow L}=2c\vec{\a}_{\sss{N-4}}=2c(\hat{e}_{\sss{N-4}}-\hat{e}_{\sss{N-3}}),\nn\\
&\lra{H \leftarrow P}=2c\a_{\sss{N-4}}+2c\a_{\sss{N-3}}=2c(\hat{e}_{\sss{N-4}}-\hat{e}_{\sss{N-2}}), \nn\\
&\lra{H \leftarrow Q}=2c\a_{\sss{N-4}}+2c\a_{\sss{N-3}}+2c\a_{\sss{N-2}}=2c(\hat{e}_{\sss{N-4}}-\hat{e}_{\sss{N-1}}),\nn\\
&\lra{H \leftarrow R}=2c\a_{\sss{N-4}}+2c\a_{\sss{N-3}}+2c\a_{\sss{N-2}}+2c\a_{\sss{N-1}}=2c(\hat{e}_{\sss{N-4}}-\hat{e}_{\sss{N}}).
\end{align}
\begin{align}
&\lra{K \leftarrow L}=\lra{M \leftarrow P}=2c\vec{\a}_{\sss{N}}=2c(\hat{e}_{\sss{N-1}}+ \hat{e}_{\sss{N}}), \nn\\
&\lra{K \leftarrow M}=\lra{L \leftarrow P}=2c\vec{\a}_{\sss{N-3}}=2c(\hat{e}_{\sss{N-3}}-\hat{e}_{\sss{N-2}}), \nn\\
&\lra{K \leftarrow P}=0, \nn\\
&\lra{K \leftarrow Q}=2c\a_{\sss{N-3}}+2c\a_{\sss{N-2}}+2c\a_{\sss{N}}=2c(\hat{e}_{\sss{N-3}} + \hat{e}_{\sss{N}}), \nn\\
&\lra{K \leftarrow R}=2c\a_{\sss{N-3}}+2c\a_{\sss{N-2}}+2c\a_{\sss{N-1}}+2c\a_{\sss{N}}=2c(\hat{e}_{\sss{N-3}} + \hat{e}_{\sss{N-1}}).
\end{align}
\begin{align}
&\lra{L \leftarrow M}=0, \nn\\
&\lra{L \leftarrow Q}=2c\vec{\a}_{\sss{N-3}}+2c\vec{\a}_{\sss{N-2}}=2c(\hat{e}_{\sss{N-3}}-\hat{e}_{\sss{N-1}}), \nn\\
&\lra{L \leftarrow R}=2c\vec{\a}_{\sss{N-3}}+2c\vec{\a}_{\sss{N-2}}+2c\vec{\a}_{\sss{N-1}}=2c(\hat{e}_{\sss{N-3}}-\hat{e}_{\sss{N}}). 
\end{align}
\begin{align}
&\lra{M \leftarrow Q}=2c\vec{\a}_{\sss{N-2}}+2c\vec{\a}_{\sss{N}}=2c(\hat{e}_{\sss{N-2}}+\hat{e}_{\sss{N}}), \nn\\
&\lra{M \leftarrow R}=2c\vec{\a}_{\sss{N-2}}+2c\vec{\a}_{\sss{N-1}}+2c\vec{\a}_{\sss{N}}=2c(\hat{e}_{\sss{N-2}}+\hat{e}_{\sss{N-1}}).
\end{align}
\begin{align}
&\lra{P \leftarrow Q}=2c\vec{\a}_{\sss{N-2}}=2c(\hat{e}_{\sss{N-2}} - \hat{e}_{\sss{N-1}}), \nn\\
&\lra{P \leftarrow R}=2c\vec{\a}_{\sss{N-2}}+2c\vec{\a}_{\sss{N-1}}=2c(\hat{e}_{\sss{N-2}}-\hat{e}_{\sss{N}}).
\end{align}
\begin{align}
\lra{Q \leftarrow R}=2c\vec{\a}_{\sss{N-1}}=2c(\hat{e}_{\sss{N-1}}-\hat{e}_{\sss{N}}).
\end{align}

\begin{figure}[ht!]
	\begin{center}
		\includegraphics[width=9cm,clip]{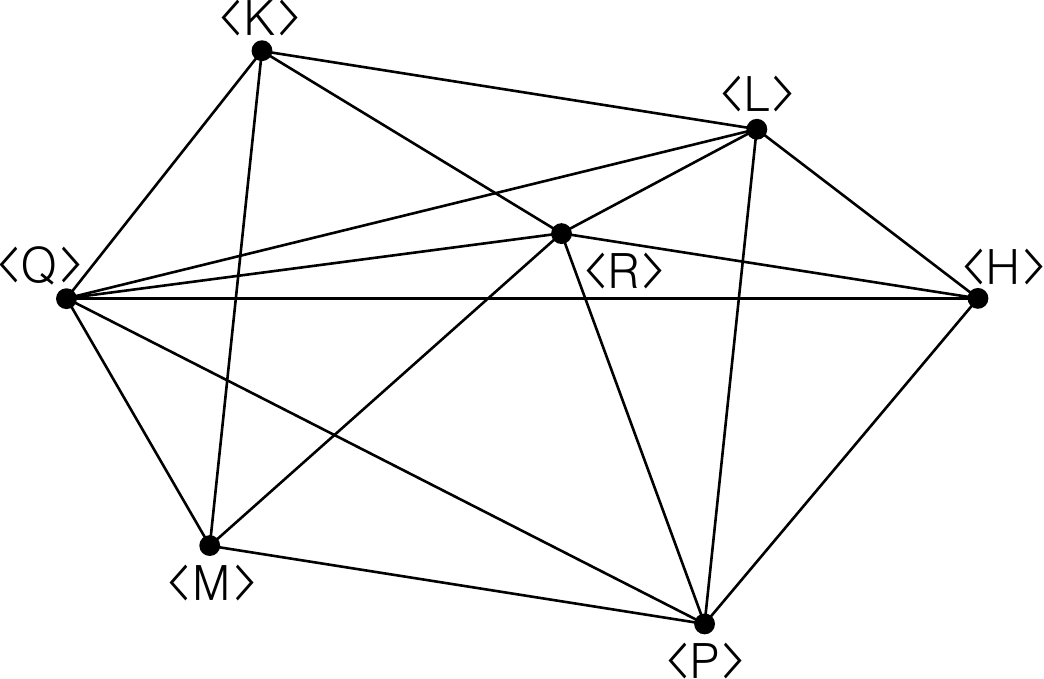}
	\end{center}
	\caption{$SO(2N)/U(N)$, $N=4m+1$. Vertices, segments and triangular faces correspond to vacua, walls and three-pronged junctions.}
	\label{fig:son_junction_4mp1}
\end{figure}

The moduli matrices for $\lra{H\leftarrow L}$, $\lra{L \leftarrow P}$ and $\lra{H \leftarrow P}$ are
\begin{align}\label{eq:single_walls_so_b}
&H_{0\lra{H\leftarrow L}}=H_{0\lra{H}} e^{e^rE_{\sss{N-4}}}, \nn\\
&H_{0\lra{L\leftarrow P}}=H_{0\lra{L}} e^{e^rE_{\sss{N-3}}}, \nn\\
&H_{0\lra{H\leftarrow P}}=H_{0\lra{H}} e^{e^r[E_{\sss{N-4}},E_{\sss{N-3}}]}.
\end{align}
The wall operators are defined by the simple root generators in (\ref{eq:gen_rt_so}):
\begin{align} \label{eq:eee2}
&E_{\sss{N-4}}=e_{\sss{N-4,N-3}}-e_{\sss{2N-3,2N-4}}, \nn\\
&E_{\sss{N-3}}=e_{\sss{N-3,N-2}}-e_{\sss{2N-2,2N-3}}, \nn\\
&\lt[ E_{\sss{N-4}},E_{\sss{N-3}}\rt]=e_{\sss{N-4,N-2}}-e_{\sss{2N-2,2N-4}}.
\end{align}

We build three-pronged junctions. We set $M_2\neq 0$, $\S_2\neq 0$ and reformulate the relevant part of the diagram in Figure \ref{fig:son_wall_4m+1} to yield the diagram in Figure \ref{fig:son_junction_4mp1}. Vertices, segments and triangular faces correspond to vacua, walls and three-pronged junctions. 

We study the three-pronged junction that divides the vacua $\{\lra{H},\lra{L},\lra{P}\}$. We apply the worldvolume symmetry transformation to the single walls in (\ref{eq:single_walls_so_b}) to produce generic wall solutions:
\begin{align}\label{eq:single_walls_wvs_so_b}
&H_{0\lra{H \leftrightarrow L}}^{({\sss{N-4}},{\sss{N-4}})}=H_{0\lra{H \leftrightarrow L}}^{({\sss{N-3}},{\sss{2N-3}})}=\exp\lt(a_{{\sss{N-4};\sss{N-4}}}+ib_{{\sss{N-4};\sss{N-4}}}\rt),\nn\\
&H_{0\lra{H \leftrightarrow L}}^{({\sss{N-4}},{\sss{N-3}})}=-H_{0\lra{H \leftrightarrow L}}^{({\sss{N-3}},{\sss{2N-4}})}=\exp\lt(a_{{\sss{N-4};\sss{N-3}}}+ib_{{\sss{N-4};\sss{N-3}}}\rt),\\
& ~ \nn\\
&H_{0\lra{L \leftrightarrow P}}^{({\sss{N-3}},{\sss{N-3}})}=H_{0\lra{L \leftrightarrow P}}^{({\sss{N-2}},{\sss{2N-2}})}=\exp\lt(a_{{\sss{N-3};\sss{N-3}}}+ib_{{\sss{N-3};\sss{N-3}}}\rt),\nn\\
&H_{0\lra{L \leftrightarrow P}}^{({\sss{N-3}},{\sss{N-2}})}=-H_{0\lra{L \leftrightarrow P}}^{({\sss{N-2}},{\sss{2N-3}})}=\exp\lt(a_{{\sss{N-3};\sss{N-2}}}+ib_{{\sss{N-3};\sss{N-2}}}\rt),\\
& ~ \nn\\
&H_{0\lra{P \leftrightarrow H}}^{({\sss{N-4}},{\sss{N-4}})}=H_{0\lra{P \leftrightarrow H}}^{({\sss{N-2}},{\sss{2N-2}})}=\exp\lt(a_{{\sss{N-4};\sss{N-4}}}+ib_{{\sss{N-4};\sss{N-4}}}\rt),\nn\\
&H_{0\lra{P \leftrightarrow H}}^{({\sss{N-4}},{\sss{N-2}})}=-H_{0\lra{P \leftrightarrow H}}^{({\sss{N-2}},{\sss{2N-4}})}=\exp\lt(a_{{\sss{N-4};\sss{N-2}}}+ib_{{\sss{N-4};\sss{N-2}}}\rt).
\end{align}

The wall solutions are obtained by the relation (\ref{eq:bps_sol}). The relevant elements of the walls are presented in \ref{app:so8mp2}.

The wall positions are computed by equalising the real parts of the flavour numbers for each colour number.  The position of $\lra{H \leftrightarrow L}$ is
\begin{align}
&\mathrm{Re}\lt(\mca_{\lra{H \leftrightarrow L}}^{({\sss{N-4}},{\sss{N-4}})}\rt)
=\mathrm{Re}\lt(\mca_{\lra{H \leftrightarrow L}}^{({\sss{N-4}},{\sss{N-3}})}\rt), \nn\\
&\mathrm{Re}\lt(\mca_{\lra{H \leftrightarrow L}}^{({\sss{N-3}},{\sss{2N-4}})}\rt)
=\mathrm{Re}\lt(\mca_{\lra{H \leftrightarrow L}}^{({\sss{N-3}},{\sss{2N-3}})}\rt),\\
& \lt(m_{\sss{N-4}}- m_{\sss{N-3}}\rt)x^1+\lt(n_{\sss{N-4}} - n_{\sss{N-3}}\rt)x^2 +
\lt(a_{\sss{N-4;N-4}}-a_{\sss{N-4;N-3}}\rt)=0. \label{eq:pos_hl_so}
\end{align}
One of the parameters $a_{\sss{N-4:I}}$, $(I=N-4,N-3)$ is independent.

The position of $\lra{L \leftrightarrow P}$ is
\begin{align}
&\mathrm{Re}\lt(\mca_{\lra{L \leftrightarrow P}}^{({\sss{N-3}},{\sss{N-3}})}\rt)
=\mathrm{Re}\lt(\mca_{\lra{L \leftrightarrow P}}^{({\sss{N-3}},{\sss{N-2}})}\rt), \nn\\
&\mathrm{Re}\lt(\mca_{\lra{L \leftrightarrow P}}^{({\sss{N-2}},{\sss{2N-3}})}\rt)
=\mathrm{Re}\lt(\mca_{\lra{L \leftrightarrow P}}^{({\sss{N-2}},{\sss{2N-2}})}\rt),\\
& \lt(m_{\sss{N-3}}- m_{\sss{N-2}}\rt)x^1+\lt(n_{\sss{N-3}} - n_{\sss{N-2}}\rt)x^2 +
\lt(a_{\sss{N-3;N-3}}-a_{\sss{N-3;N-2}}\rt)=0.  \label{eq:pos_lp_so}
\end{align}
One of the parameters $a_{\sss{N-4:I}}$, $(I=N-3,N-2)$ is independent.

The position of $\lra{P \leftrightarrow H}$ is
\begin{align}
&\mathrm{Re}\lt(\mca_{\lra{P \leftrightarrow H}}^{({\sss{N-4}},{\sss{N-4}})}\rt)
=\mathrm{Re}\lt(\mca_{\lra{P \leftrightarrow H}}^{({\sss{N-4}},{\sss{N-2}})}\rt), \nn\\
&\mathrm{Re}\lt(\mca_{\lra{P \leftrightarrow H}}^{({\sss{N-2}},{\sss{2N-4}})}\rt)
=\mathrm{Re}\lt(\mca_{\lra{P \leftrightarrow H}}^{({\sss{N-2}},{\sss{2N-2}})}\rt),\\
& \lt(m_{\sss{N-4}}- m_{\sss{N-2}}\rt)x^1+\lt(n_{\sss{N-4}} - n_{\sss{N-2}}\rt)x^2 +
\lt(a_{\sss{N-4;N-4}}-a_{\sss{N-4;N-2}}\rt)=0.\label{eq:pos_ph_so}
\end{align}
One of the parameters $a_{\sss{N-4:I}}$, $(I=N-4,N-2)$ is independent.

There is a consistency condition:
\begin{align}\label{eq:so_hlp_cond}
(a_{\sss{N-4;N-3}}-a_{\sss{N-4;N-2}})-(a_{\sss{N-3;N-3}}-a_{\sss{N-3;N-2}})=0.
\end{align}
Therefore two independent parameters describe the junction position. The position of the junction that divides $\lt\{\lra{H},\lra{L},\lra{P}\rt\}$ is
\begin{align}\label{eq:so_hlp_pos}
&(x,y)=\lt(\frac{U_1}{U_3},\frac{U_2}{U_3}\rt), \nn\\
&U_1=(a_{\sss{N-3;N-3}}-a_{\sss{N-3;N-2}})(n_{\sss{N-4}}-n_{\sss{N-3}}) \nn\\
&\quad\quad -(a_{\sss{N-4;N-4}}-a_{\sss{N-4;N-3}})(n_{\sss{N-3}}-n_{\sss{N-2}}), \nn\\
&U_2=-(a_{\sss{N-3;N-3}}-a_{\sss{N-3;N-2}})(m_{\sss{N-4}}-m_{\sss{N-3}}) \nn\\
&\quad\quad +(a_{\sss{N-4;N-4}}-a_{\sss{N-4;N-3}})(m_{\sss{N-3}}-m_{\sss{N-2}}), \nn\\
&U_3=m_{\sss{N-4}}(n_{\sss{N-3}}-n_{\sss{N-2}})+m_{\sss{N-3}}(-n_{\sss{N-4}}+n_{\sss{N-2}})
+m_{\sss{N-2}}(n_{\sss{N-4}}-n_{\sss{N-3}}).
\end{align}
The solution (\ref{eq:so_hlp_pos}) is constrained by (\ref{eq:so_hlp_cond}).

\section{Three-pronged junctions of mass-deformed nonlinear sigma models on $Sp(N)/U(N)$} \label{sec:sp}
\setcounter{equation}{0}

\begin{figure}[ht!]
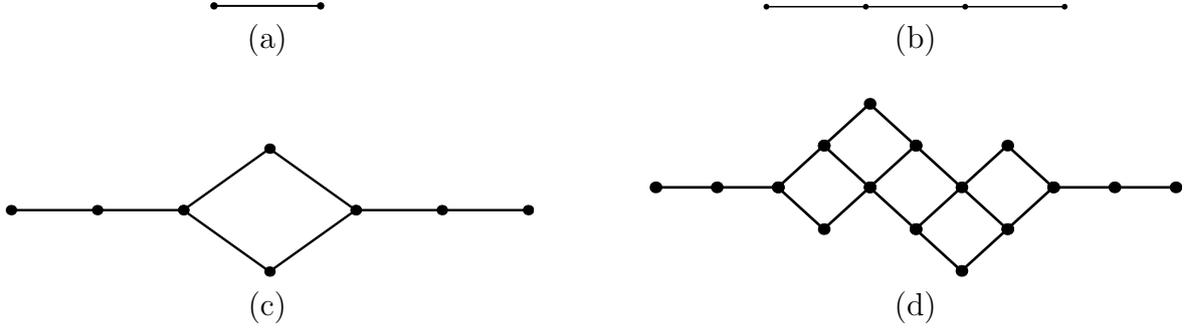

\begin{center}
$\begin{array}{ccc}
\includegraphics[width=1.5cm,clip]{line1.pdf}
&~~~~~~&
\includegraphics[width=4cm,clip]{line2.pdf}\\
\mathrm{(a)} &~~~~~~& \mathrm{(b)} \\
~~ &  ~~ & ~~ \\
\includegraphics[width=7cm,clip]{diamond.pdf}
&~~~~~~&
\includegraphics[width=7cm,clip]{sim.pdf}\\
\mathrm{(c)} &~~~~~~& \mathrm{(d)}
\end{array}
$
\end{center}
 \caption{There is a recurrence of one of the  diagrams for each $N$ mod 4. (a)$N=4m-3$, (b)$N=4m-2$, (c)$N=4m-1$, and (d)$N=4m$, $(m\geq 1)$. }
 \label{fig:spn}
\end{figure}

Vacua and elementary walls of mass-deformed nonlinear sigma models on $Sp(N)/U(N)$ are studied in the pictorial representation \cite{Arai:2018tkf}. All the diagrams of vacua and elementary walls are symmetric. There is a recurrence of a diagram for each $N$ mod 4 in the vacuum structures that are connected to the maximum number of elementary walls. The vacuum structures are proved by induction. The diagrams are presented in Figure \ref{fig:spn}. The whole structure of vacua and elementary walls of the mass-deformed nonlinear sigma models on $Sp(N)/U(N)$ for generic $N$ can be derived from the vacuum structures that are connected to the maximum number of elementary walls.

In this section we construct three-pronged junctions from the diagrams (c) and (d) of Figure \ref{fig:spn} for generic $N$.

The vacua are labelled in the descending order \cite{Arai:2018tkf}:
\bea
&&(\S_1,\S_2,\cdots,\S_{N-1},\S_N)=(m_1,m_2,\cdots,m_{N-1},m_N), \nn\\
&&(\S_1,\S_2,\cdots,\S_{N-1},\S_N)=(m_1,m_2,\cdots,m_{N-1},-m_N), \nn\\
&&(\S_1,\S_2,\cdots,\S_{N-1},\S_N)=(m_1,m_2,\cdots,-m_{N-1},m_N), \nn\\
&&(\S_1,\S_2,\cdots,\S_{N-1},\S_N)=(m_1,m_2,\cdots,-m_{N-1},-m_N), \nn\\
&& \quad \vdots \nn\\
&&(\S_1,\S_2,\cdots,\S_{N-1},\S_N)=(m_1,-m_2,\cdots,-m_{N-1},-m_N), \nn\\
&&(\S_1,\S_2,\cdots,\S_{N-1},\S_N)=(-m_1,m_2,\cdots,m_{N-1},m_N), \nn\\
&&\quad \vdots \nn\\
&&(\S_1,\S_2,\cdots,\S_{N-1},\S_N)=(-m_1,-m_2,\cdots,-m_{N-1},-m_N).
\eea

\subsection{$N=4m-1$}  \label{sec:sp_n=4m-1}

\begin{figure}[ht!]
\vspace{1cm}
\begin{center}
\includegraphics[width=7cm,clip]{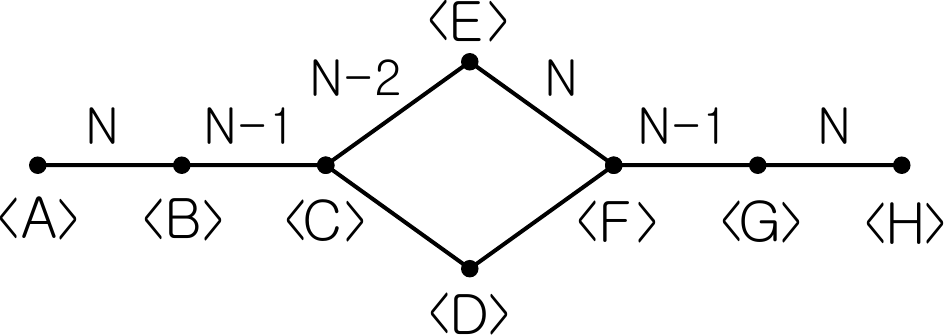}
\end{center}
 \caption{ Vacua and elementary walls of the mass-deformed nonlinear sigma model on $Sp(N)/U(N)$, $N=4m-1$ \cite{Lee:2017kaj}. The numbers in brackets indicate the vacuum labels. The numbers without brackets indicate the subscript $i$'s of simple roots $\vec{\a}_i$, $(i=N-2,\cdots,N)$.}
 \label{fig:spn_wall_4m-1}
\end{figure}

In this subsection, we construct three-pronged junctions of the mass-deformed nonlinear sigma model on $Sp(N)/U(N)$ with $N=4m-1$, $(m\geq 1)$. We reduce the model (\ref{eq:mainlag}) by setting $M_2=0$ and $\S_2=0$ to investigate the structure of vacua and elementary walls. The diagram in Figure \ref{fig:spn_wall_4m-1} recurs at the vacua that are connected to the maximum number of elementary walls \cite{Arai:2018tkf}.
Elementary walls are identified with scaled simple roots $\vec{g}_{\lra{\cdot \leftarrow \cdot}}=2c\vec{\a}_i$, $(i=1,\cdots,N-1)$ and $\vec{g}_{\lra{\cdot\leftarrow\cdot}}=c\vec{\a}_{\sss{N}}$. Compressed walls are linear combination of the scaled simple roots \cite{Arai:2018tkf,Kim:2019jzo}. The single walls of the recurring diagram are derived below:
\begin{align}
&\lra{A\leftarrow B}=\lra{C\leftarrow D}=\lra{E\leftarrow F}=\lra{G\leftarrow H}= c\vec{\a}_{\sss{N}}=2c\hat{e}_{\sss{N}}, \nn\\
&\lra{A\leftarrow C}=\lra{B\leftarrow D}= \lra{E\leftarrow G}=\lra{F\leftarrow H}= 2c\vec{\a}_{\sss{N-1}}+c\vec{\a}_{\sss{N}}=2c\hat{e}_{\sss{N-1}},\nn\\
&\lra{A\leftarrow D}=\lra{E\leftarrow H}= 0,\nn\\
&\lra{A\leftarrow E}=\lra{B\leftarrow F}= \lra{C\leftarrow G}=\lra{D\leftarrow H}= 2c\vec{\a}_{\sss{N-2}}+2c\vec{\a}_{\sss{N-1}}+c\vec{\a}_{\sss{N}}=2c\hat{e}_{\sss{N-2}},\nn\\
&\lra{A\leftarrow F}=\lra{C\leftarrow H}= 0,\nn\\
&\lra{A\leftarrow G}=\lra{B\leftarrow H}= 0,\nn\\
&\lra{A\leftarrow H}= 0.
\end{align}
\begin{align}
&\lra{B\leftarrow C}=\lra{F\leftarrow G}= 2c\vec{\a}_{\sss{N-1}}=2c( \hat{e}_{\sss{N-1}} -  \hat{e}_{\sss{N}}),\nn\\ 
&\lra{B\leftarrow E}= \lra{D\leftarrow G}= 2c\vec{\a}_{\sss{N-2}}+2c\vec{\a}_{\sss{N-1}}=2c( \hat{e}_{\sss{N-2}} -  \hat{e}_{\sss{N}}), \nn\\
&\lra{B\leftarrow G}= 0, \nn\\
&\lra{C\leftarrow E}=\lra{D\leftarrow F}= 2c\vec{\a}_{\sss{N-2}}=2c(\hat{e}_{\sss{N-2}}-\hat{e}_{\sss{N-1}}),\nn\\
&\lra{C\leftarrow F}= 0,\nn\\
&\lra{D\leftarrow E}= 0.\nn\\
\end{align}

\begin{figure}[ht!]
\begin{center}
\includegraphics[width=9cm,clip]{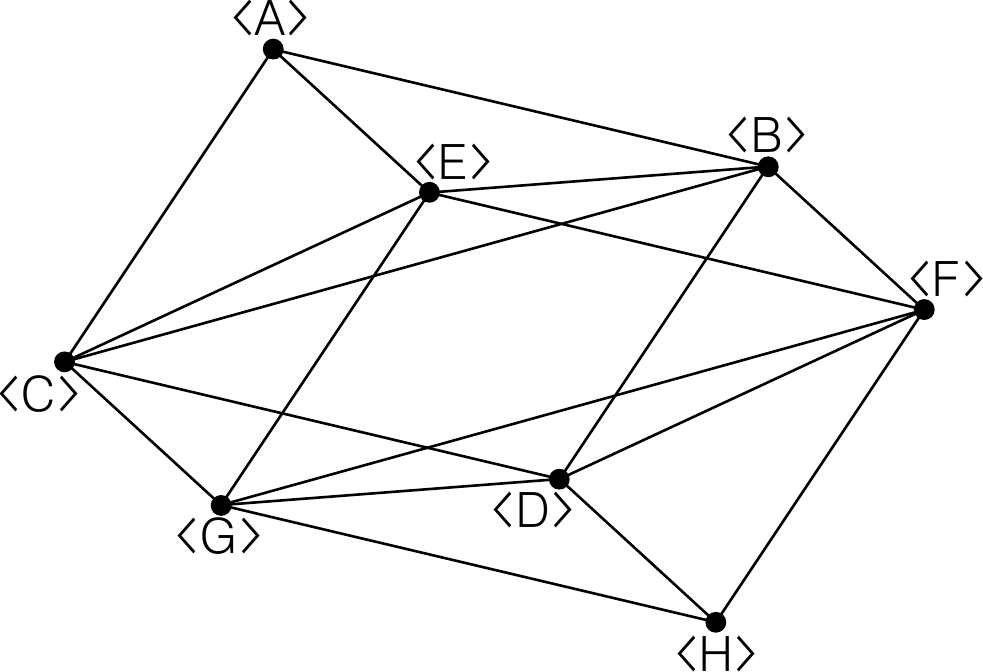}
\end{center}
 \caption{$Sp(N)/U(N)$, $N=4m-1$. Vertices, segments and triangular faces correspond to vacua, walls and three-pronged junctions.}
 \label{fig:spn_junction_4m-1}
\end{figure}

The moduli matrices for $\lra{A\leftarrow B}$, $\lra{B\leftarrow C}$ and $\lra{A\leftarrow C}$ can be read from the diagram in Figure \ref{fig:spn_wall_4m-1}:
\begin{align} \label{eq:single_walls_sp_a}
&H_{0\lra{A\leftarrow B}}=H_{0\lra{A}} e^{e^rE_{\sss{N}}}, \nn\\
&H_{0\lra{B\leftarrow C}}=H_{0\lra{B}} e^{e^rE_{\sss{N-1}}}, \nn\\
&H_{0\lra{A\leftarrow C}}=H_{0\lra{A}} e^{e^r[[E_{\sss{N}},E_{\sss{N-1}}],E_{\sss{N-1}}]}.
\end{align}
The wall operators are defined by the simple root generators in (\ref{eq:gen_rt_sp}):
\begin{align}\label{eq:eee3}
&E_{\sss{N}}=e_{\sss{N,2N}}, \nn\\
&E_{\sss{N-1}}=e_{\sss{N-1,N}}-e_{\sss{2N,2N-1}}, \nn\\
&\lt[\lt[ E_{\sss{N}},E_{\sss{N-1}}\rt],E_{\sss{N-1}}\rt]=2e_{\sss{N-1,2N-1}}.
\end{align}
The wall operator of $\lra{A\leftarrow C}$ is defined by a multiple commutator since the sector involves the summation of unequal length vectors.

We construct three-pronged junctions. We set $M_2\neq 0$ and $\S_2\neq 0$ and reformulate the diagram in Figure \ref{fig:spn_wall_4m-1} to yield the diagram in Figure \ref{fig:spn_junction_4m-1}. Vertices, segments and triangular faces correspond to vacua, walls and three-pronged junctions.

We study the three-pronged junction that divides the vacua $\{\lra{A},\lra{B},\lra{C}\}$. We transform the moduli matrices in (\ref{eq:single_walls_sp_a}) to produce generic wall solutions. The relevant elements of the moduli matrices for $\{\lra{A \leftrightarrow B},\lra{B \leftrightarrow C},\lra{C \leftrightarrow A}\}$ are listed below:
\begin{align} \label{eq:single_walls_wvs_sp_a}
&H_{0\lra{A \leftrightarrow B}}^{({\sss{N}},{\sss{N}})}=\exp\lt(a_{{\sss{N};\sss{N}}}+ib_{{\sss{N};\sss{N}}}\rt),\nn\\
&H_{0\lra{A \leftrightarrow B}}^{({\sss{N}},{\sss{2N}})}=\exp\lt(a_{{\sss{N};\sss{2N}}}+ib_{{\sss{N};\sss{2N}}}\rt). \nn\\
& ~\nn\\
&H_{0\lra{B \leftrightarrow C}}^{({\sss{N-1}},{\sss{N-1}})}=H_{0\lra{B \leftrightarrow C}}^{({\sss{N}},{\sss{2N}})}=\exp\lt(a_{{\sss{N-1};\sss{N-1}}}+ib_{{\sss{N-1};\sss{N-1}}}\rt),\nn\\
&H_{0\lra{B \leftrightarrow C}}^{({\sss{N-1}},{\sss{N}})}=-H_{0\lra{B \leftrightarrow C}}^{({\sss{N}},{\sss{2N-1}})}=\exp\lt(a_{{\sss{N-1};\sss{N}}}+ib_{{\sss{N-1};\sss{N}}}\rt). \nn\\
&~\nn\\
&H_{0\lra{C \leftrightarrow A}}^{({\sss{N-1}},{\sss{N-1}})}=\exp\lt(a_{{\sss{N-1};\sss{N-1}}}+ib_{{\sss{N-1};\sss{N-1}}}\rt),\nn\\
&H_{0\lra{C \leftrightarrow A}}^{({\sss{N-1}},{\sss{2N-1}})}=\exp\lt(a_{{\sss{N-1};\sss{2N-1}}}+ib_{{\sss{N-1};\sss{2N-1}}}\rt).
\end{align}
The first superscript is the row number and the second superscript is the column number.

The wall solutions are obtained by the relation (\ref{eq:bps_sol}). The relevant elements of the walls are presented in \ref{app:sp4mm1}.

The walls are located at the positions where the real parts of the elements in the same row are equal. The position of $\lra{A\leftrightarrow B}$ is
\begin{align}\label{eq:pos_ab_sp}
&\mathrm{Re}\lt(\mca_{\lra{A \leftrightarrow B}}^{({\sss{N}},{\sss{N}})}\rt)=
\mathrm{Re}\lt(\mca_{\lra{A \leftrightarrow B}}^{({\sss{N}},{\sss{2N}})}\rt), \nn\\
&2m_{\sss{N}}x^1+2n_{\sss{N}}x^2+\lt(a_{\sss{N;N}}-a_{\sss{N;2N}}\rt)=0.
\end{align}
Only one of the parameters $a_{\sss{N;I}}$, $(I=N,2N)$ is independent as the moduli matrices in (\ref{eq:single_walls_wvs_sp_a}) have the worldvolume symmetry (\ref{eq:wvs}).

The position of $\lra{B \leftrightarrow C}$ is
\begin{align}\label{eq:pos_bc_sp}
&\mathrm{Re}\lt(\mca_{\lra{B \leftrightarrow C}}^{({\sss{N-1}},{\sss{N-1}})}\rt)=
\mathrm{Re}\lt(\mca_{\lra{B \leftrightarrow C}}^{({\sss{N-1}},{\sss{N}})}\rt), \nn\\
&\mathrm{Re}\lt(\mca_{\lra{B \leftrightarrow C}}^{({\sss{N}},{\sss{2N-1}})}\rt)=
\mathrm{Re}\lt(\mca_{\lra{B \leftrightarrow C}}^{({\sss{N}},{\sss{2N}})}\rt), \nn\\
&\lt(m_{\sss{N-1}}-m_{\sss{N}}\rt)x^1+\lt(n_{\sss{N-1}}-n_{\sss{N}}\rt)x^2+\lt(a_{\sss{N-1;N-1}}-a_{\sss{N-1;N}}\rt)=0.
\end{align}
Only one of the parameters $a_{\sss{N-1;I}}$, $(I=N-1,N)$ is independent.

The position of $\lra{C \leftrightarrow A}$ is
\begin{align}\label{eq:pos_ca_sp}
&\mathrm{Re}\lt(\mca_{\lra{C \leftrightarrow A}}^{({\sss{N-1}},{\sss{N-1}})}\rt)=
\mathrm{Re}\lt(\mca_{\lra{C \leftrightarrow A}}^{({\sss{N-1}},{\sss{2N-1}})}\rt), \nn\\
&2m_{\sss{N-1}}x^1+ 2n_{\sss{N-1}}x^2+\lt(a_{\sss{N-1;N-1}}-a_{\sss{N-1;2N-1}}\rt)=0.
\end{align}
Only one of the parameters $a_{\sss{N-1;I}}$, $(I=N-1,2N-1)$ is independent.

There is a consistency condition:
\begin{align}\label{eq:sp_abc_cond}
\lt(a_{\sss{N-1;N-1}}-a_{\sss{N-1;2N-1}}\rt)-\lt(a_{\sss{N;N}}-a_{\sss{N;2N}}\rt)
-2\lt(a_{\sss{N-1;N-1}}-a_{\sss{N-1;N}}\rt)=0.
\end{align} 
Therefore there are only two independent parameters to specify the junction position. The position of the junction that divides $\{\lra{A},\lra{B},\lra{C}\}$ is calculated from (\ref{eq:pos_ab_sp}), (\ref{eq:pos_bc_sp}) and (\ref{eq:pos_ca_sp}):
\begin{align}\label{eq:sp_abc_pos}
&(x,y)=\lt(\frac{T_1}{T_3},\frac{T_2}{T_3}\rt),\nn\\
&T_1=-2(a_{\sss{N;N}}-a_{\sss{N;2N}})n_{\sss{N-1}}+2(a_{\sss{N-1;N-1}}-a_{\sss{N-1;2N-1}})n_{\sss{N}},\nn\\
&T_2=2(a_{\sss{N;N}}-a_{\sss{N;2N}})m_{\sss{N-1}} -2(a_{\sss{N-1;N-1}}-a_{\sss{N-1;2N-1}})m_{\sss{N}}, \nn\\
&T_3=-4m_{\sss{N-1}}n_{\sss{N}}+4m_{\sss{N}}n_{\sss{N-1}}.
\end{align}
The solution (\ref{eq:sp_abc_pos}) is constrained by (\ref{eq:sp_abc_cond}).

\subsection{$N=4m$} \label{sec:sp_n=4m}

In this subsection, we construct three-pronged junctions of the mass-deformed nonlinear sigma model on $Sp(N)/U(N)$ with $N=4m$. The structure of vacua and walls can be analysed in the reduced mass-deformed model with $M_2=0$ and $\S_2=0$. There is a recurrence of the diagram in Figure \ref{fig:spn_wall_4m} at the vacua that are connected to the maximum number of elementary walls \cite{Arai:2018tkf}. The single walls that interpolate the vacua $\{\lra{H}$,$\lra{K}$,$\lra{L}$,$\lra{M}$,$\lra{P}$,$\lra{Q}$,$\lra{R}\}$ are derived from the diagram in Figure \ref{fig:spn_wall_4m}:
\begin{figure}[ht!]
\begin{center}
\includegraphics[width=10cm,clip]{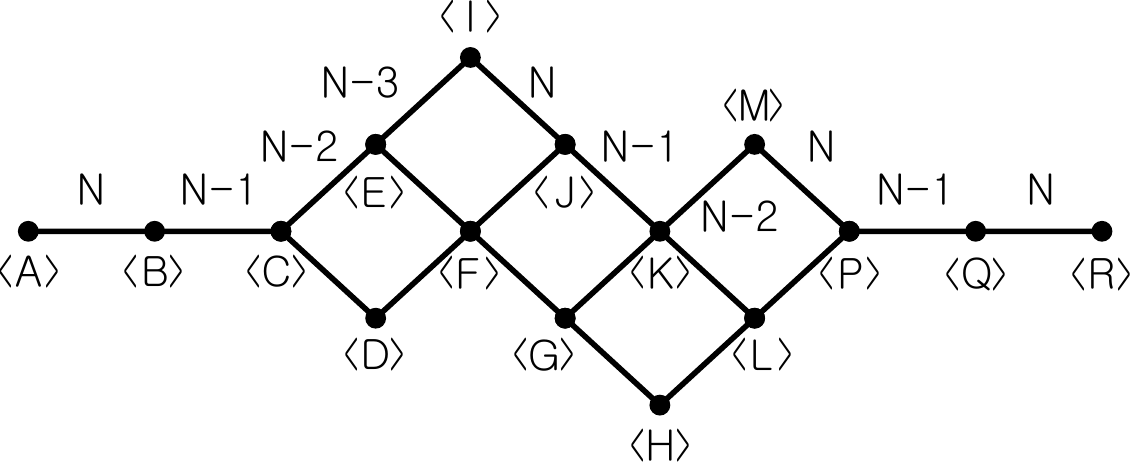}
\end{center}
 \caption{ Vacua and elementary walls of the mass-deformed nonlinear sigma model on $Sp(N)/U(N)$, $N=4m$ \cite{Lee:2017kaj}. The numbers in brackets indicate the vacuum labels. The numbers without brackets indicate the subscript $i$'s of simple roots $\vec{\a}_i$, $(i=N-3,\cdots,N)$.}
 \label{fig:spn_wall_4m}
\end{figure}

\begin{align}
&\lra{H\leftarrow K}=0, \nn\\
&\lra{H\leftarrow L}=2c\vec{\a}_{\sss{N-3}}=2c(\hat{e}_{\sss{N-3}}-\hat{e}_{\sss{N-2}}), \nn\\
&\lra{H\leftarrow M}=0, \nn\\
&\lra{H\leftarrow P}=2c\vec{\a}_{\sss{N-3}}+2c\vec{\a}_{\sss{N-2}}=2c(\hat{e}_{\sss{N-3}}-\hat{e}_{\sss{N-1}}), \nn\\
&\lra{H\leftarrow Q}=2c\vec{\a}_{\sss{N-3}}+2c\vec{\a}_{\sss{N-2}}+2c\vec{\a}_{\sss{N-1}}=2c(\hat{e}_{\sss{N-3}}-\hat{e}_{\sss{N}}), \nn\\
&\lra{H\leftarrow R}=2c\vec{\a}_{\sss{N-3}}+2c\vec{\a}_{\sss{N-2}}+2c\vec{\a}_{\sss{N-1}}+c\vec{\a}_{\sss{N}}=2c\hat{e}_{\sss{N-3}}.
\end{align}
\begin{align}
&\lra{K \leftarrow L}=\lra{M \leftarrow P}=\lra{Q \leftarrow R}=c\vec{\a}_{\sss{N}}=2c\hat{e}_{\sss{N}}, \nn\\
&\lra{K \leftarrow M}=\lra{L \leftarrow P}=2c\vec{\a}_{\sss{N-2}}=2c(\hat{e}_{\sss{N-2}}-\hat{e}_{\sss{N-1}}), \nn\\
&\lra{K \leftarrow P}=0, \nn\\
&\lra{K \leftarrow Q}=\lra{L \leftarrow R}=2c\vec{\a}_{\sss{N-2}}+2c\vec{\a}_{\sss{N-1}}+c\vec{\a}_{\sss{N}}=2c\hat{e}_{\sss{N-2}}, \nn\\
&\lra{K \leftarrow R}=0.
\end{align}
\begin{align}
&\lra{L \leftarrow M}=0, \nn\\
&\lra{L \leftarrow Q}=2c\vec{\a}_{\sss{N-2}}+2c\vec{\a}_{\sss{N-1}}=2c(\hat{e}_{\sss{N-2}}-\hat{e}_{\sss{N}}), \nn\\
&\lra{M \leftarrow Q}=\lra{P \leftarrow R}=2c\vec{\a}_{\sss{N-1}}+c\vec{\a}_{\sss{N}}=2c\hat{e}_{\sss{N-1}}, \nn\\
&\lra{M \leftarrow R}=0, \nn \\
&\lra{P \leftarrow Q}=2c\vec{\a}_{\sss{N-1}}=2c(\hat{e}_{\sss{N-1}}-\hat{e}_{\sss{N}}).
\end{align}

\begin{figure}[ht!]
	\begin{center}
		\includegraphics[width=9cm,clip]{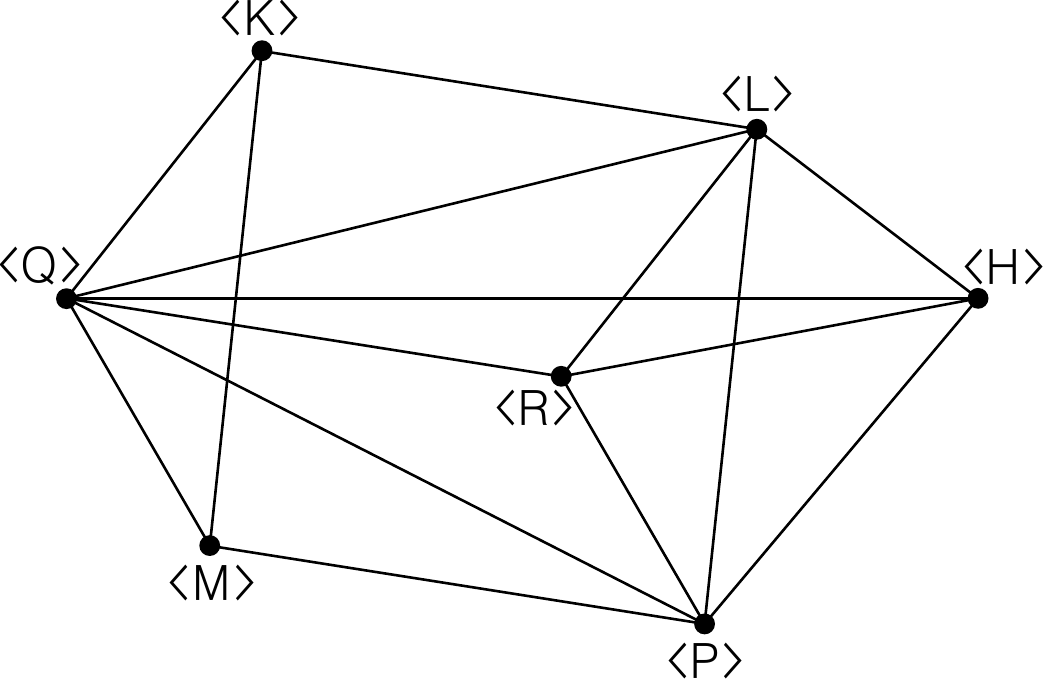}
	\end{center}
	\caption{$Sp(N)/U(N)$, $N=4m$. Vertices, segments and triangular faces correspond to vacua, walls and three-pronged junctions.}
	\label{fig:spn_junction_4m}
\end{figure}

The moduli matrices for $\lra{H\leftarrow L}$, $\lra{L\leftarrow P}$ and $\lra{H\leftarrow P}$ are
\begin{align} \label{eq:single_walls_sp_b}
&H_{0\lra{H\leftarrow L}}=H_{0\lra{H}}e^{e^rE_{N-3}}, \nn\\
&H_{0\lra{L\leftarrow P}}=H_{0\lra{L}}e^{e^rE_{N-2}}, \nn\\
&H_{0\lra{H\leftarrow P}}=H_{0\lra{H}}e^{e^r[E_{N-3},E_{N-2}]}.
\end{align}
The wall operators are defined by the simple root generators in (\ref{eq:gen_rt_sp}):
\begin{align}\label{eq:eee4}
&E_{\sss{N-3}}=e_{\sss{N-3,N-2}}-e_{\sss{2N-2,2N-3}}, \nn\\
&E_{\sss{N-2}}=e_{\sss{N-2,N-1}}-e_{\sss{2N-1,2N-2}}, \nn\\
&\lt[ E_{\sss{N-3}},E_{\sss{N-2}}\rt]=e_{\sss{N-3,N-1}} - e_{\sss{2N-1,2N-3}}.
\end{align}

We build three-pronged junctions. We set $M_2\neq 0$, $\S_2\neq 0$ and reformulate the relevant part of the diagram in Figure \ref{fig:spn_wall_4m} to yield the diagram in Figure \ref{fig:spn_junction_4m}. Vertices, segments and triangular faces correspond to vacua, walls and three-pronged junctions.

We study the three-pronged junction that divides the vacua $\{\lra{H},\lra{L},\lra{P}\}$. We apply the worldvolume symmetry transformation to the single walls in (\ref{eq:single_walls_sp_b}) to produce generic wall solutions:
\begin{align}\label{eq:single_walls_wvs_sp_b}
&H_{0\lra{H\leftrightarrow L}}^{({\sss{N-3}},{\sss{N-3}})}=H_{0\lra{H\leftrightarrow L}}^{({\sss{N-2}},{\sss{2N-2}})}=\exp(a_{\sss{N-3;N-3}}+ib_{\sss{N-3;N-3}}), \nn\\
&H_{0\lra{H\leftrightarrow L}}^{({\sss{N-3}},{\sss{N-2}})}=-H_{0\lra{H\leftrightarrow L}}^{({\sss{N-2}},{\sss{2N-3}})}=\exp(a_{\sss{N-3;N-2}}+ib_{\sss{N-3;N-2}}),\\
&~\nn\\
&H_{0\lra{L\leftrightarrow P}}^{({\sss{N-2}},{\sss{N-2}})}=H_{0\lra{L\leftrightarrow P}}^{({\sss{N-1}},{\sss{2N-1}})}=\exp(a_{\sss{N-2;N-2}}+ib_{\sss{N-2;N-2}}), \nn\\
&H_{0\lra{L\leftrightarrow P}}^{({\sss{N-2}},{\sss{N-1}})}=-H_{0\lra{L\leftrightarrow P}}^{({\sss{N-1}},{\sss{2N-2}})}=\exp(a_{\sss{N-2;N-1}}+ib_{\sss{N-2;N-1}}),\\
&~\nn\\
&H_{0\lra{P\leftrightarrow H}}^{({\sss{N-3}},{\sss{N-3}})}=H_{0\lra{P\leftrightarrow H}}^{({\sss{N-1}},{\sss{2N-1}})}=\exp(a_{\sss{N-3;N-3}}+ib_{\sss{N-3;N-3}}), \nn\\
&H_{0\lra{P\leftrightarrow H}}^{({\sss{N-3}},{\sss{N-1}})}=-H_{0\lra{P\leftrightarrow H}}^{({\sss{N-1}},{\sss{2N-3}})}=\exp(a_{\sss{N-3;N-1}}+ib_{\sss{N-3;N-1}}).
\end{align}

The wall solutions are obtained by the relation (\ref{eq:bps_sol}). The relevant elements of the walls are presented in \ref{app:sp4m}.

The wall positions are computed by equalising the real parts of the flavour numbers for each colour number. The position of $\lra{H \leftrightarrow L}$ is
\begin{align}
&\mathrm{Re}\lt(\mca_{\lra{H \leftrightarrow L}}^{({\sss{N-3}},{\sss{N-3}})}\rt)
=\mathrm{Re}\lt(\mca_{\lra{H \leftrightarrow L}}^{({\sss{N-3}},{\sss{N-2}})}\rt), \nn\\
&\mathrm{Re}\lt(\mca_{\lra{H \leftrightarrow L}}^{({\sss{N-2}},{\sss{2N-3}})}\rt)
=\mathrm{Re}\lt(\mca_{\lra{H \leftrightarrow L}}^{({\sss{N-2}},{\sss{2N-2}})}\rt),\nn\\
& \lt(m_{\sss{N-3}}- m_{\sss{N-2}}\rt)x^1+\lt(n_{\sss{N-3}} - n_{\sss{N-2}}\rt)x^2 +
\lt(a_{\sss{N-3;N-3}}-a_{\sss{N-3;N-2}}\rt)=0.
\end{align}
One of the parameters $a_{\sss{N-3;I}}$, $(I=N-3,N-2)$ is independent.

The position of $\lra{L \leftrightarrow P}$ is
\begin{align}
&\mathrm{Re}\lt(\mca_{\lra{L \leftrightarrow P}}^{({\sss{N-2}},{\sss{N-2}})}\rt)
=\mathrm{Re}\lt(\mca_{\lra{L \leftrightarrow P}}^{({\sss{N-2}},{\sss{N-1}})}\rt), \nn\\
&\mathrm{Re}\lt(\mca_{\lra{L \leftrightarrow P}}^{({\sss{N-1}},{\sss{2N-2}})}\rt)
=\mathrm{Re}\lt(\mca_{\lra{L \leftrightarrow P}}^{({\sss{N-1}},{\sss{2N-1}})}\rt),\nn\\
& \lt(m_{\sss{N-2}}- m_{\sss{N-1}}\rt)x^1+\lt(n_{\sss{N-2}} - n_{\sss{N-1}}\rt)x^2 +
\lt(a_{\sss{N-2;N-2}}-a_{\sss{N-2;N-1}}\rt)=0.
\end{align}
One of the parameters $a_{\sss{N-2;I}}$, $(I=N-2,N-1)$ is independent.

The position of $\lra{P \leftrightarrow H}$ is
\begin{align}
&\mathrm{Re}\lt(\mca_{\lra{P \leftrightarrow H}}^{({\sss{N-3}},{\sss{N-3}})}\rt)
=\mathrm{Re}\lt(\mca_{\lra{P \leftrightarrow H}}^{({\sss{N-3}},{\sss{N-1}})}\rt), \nn\\
&\mathrm{Re}\lt(\mca_{\lra{P \leftrightarrow H}}^{({\sss{N-1}},{\sss{2N-3}})}\rt)
=\mathrm{Re}\lt(\mca_{\lra{P \leftrightarrow H}}^{({\sss{N-1}},{\sss{2N-1}})}\rt),\nn\\
& \lt(m_{\sss{N-3}}- m_{\sss{N-1}}\rt)x^1+\lt(n_{\sss{N-3}} - n_{\sss{N-1}}\rt)x^2 +
\lt(a_{\sss{N-3;N-3}}-a_{\sss{N-3;N-1}}\rt)=0.
\end{align}
One of the parameters $a_{\sss{N-3;I}}$, $(I=N-3,N-1)$ is independent.

There is a consistency condition:
\begin{align}\label{eq:sp_hlp_cond}
(a_{\sss{N-3;N-2}}-a_{\sss{N-3;N-1}})-(a_{\sss{N-2;N-2}}-a_{\sss{N-2;N-1}})=0.
\end{align}
Therefore two independent parameters describe the junction position. The position of the junction that divides $\{\lra{H},\lra{L},\lra{P}\}$ is
\begin{align}\label{eq:sp_hlp_pos}
&(x,y)=\lt(\frac{V_1}{V_3},\frac{V_2}{V_3}\rt), \nn\\
&V_1=(a_{\sss{N-2;N-2}} - a_{\sss{N-2;N-1}})(n_{\sss{N-3}}-n_{\sss{N-2}}) \nn\\
&\quad\quad -(a_{\sss{N-3;N-3}} - a_{\sss{N-3;N-2}})(n_{\sss{N-2}}-n_{\sss{N-1}}), \nn\\
&V_2=-(a_{\sss{N-2;N-2}} - a_{\sss{N-2;N-1}})(m_{\sss{N-3}}-m_{\sss{N-2}}) \nn\\
&\quad\quad +(a_{\sss{N-3;N-3}} - a_{\sss{N-3;N-2}})(m_{\sss{N-2}}-m_{\sss{N-1}}), \nn\\
&V_3=m_{\sss{N-3}}(n_{\sss{N-2}}-n_{\sss{N-1}})+m_{\sss{N-2}}(-n_{\sss{N-3}}+n_{\sss{N-1}})
+m_{\sss{N-1}}(n_{\sss{N-3}}-n_{\sss{N-2}}).
\end{align}
The solution (\ref{eq:sp_hlp_pos}) is constrained by (\ref{eq:sp_hlp_cond}).

The sector $\{\lra{H},\lra{L},\lra{P}\}$, which consists of equal length simple roots, of the recurring diagram in Figure \ref{fig:spn_wall_4m} of the mass-deformed nonlinear sigma models on $Sp(N)/U(N)$ with $N=4m$ have the same  
structure as the sector $\{\lra{H},\lra{L},\lra{P}\}$ of the recurring diagram in Figure \ref{fig:son_wall_4m+1} of the mass-deformed nonlinear sigma models on $SO(2N)/U(N)$ with $N=4m+1$.

\section{${\mathcal{N}}=2$ nonlinear sigma models on the Grassmann manifold}
\label{sec:tastgr}
\setcounter{equation}{0}

In this section, we review and discuss the ${\mathcal{N}}=2$ nonlinear sigma models on the Grassmann manifold written in the ${\mathcal{N}}=1$ superspace formalism \cite{Lindstrom:1983rt,Rocek:1980kc}, in the harmonic superspace formalism \cite{Galperin:2001uw,Galperin:1985dw} and in the projective superspace formalism \cite{Arai:2006gg}. We show that the on-shell component Lagrangian of the mass-deformed nonlinear sigma model on the Grassmann manifold that is obtained in the harmonic superspace formalism is equivalent to the on-shell component Lagrangian that is obtained in the ${\mathcal{N}}=1$ superspace formalism. We also show that the potential of the nonlinear sigma model on the complex projective space that is obtained in the projective superspace formalism is equivalent to the potential with the FI parameters $c^a=(0,0,c=1)$ that is obtained in the ${\mathcal{N}}=1$ superspace formalism.

In Section \ref{sec:n1}, we review the ${\mathcal{N}}=2$ nonlinear sigma model on the Grassmann manifold obtained in the ${\mathcal{N}}=1$ superspace formalism. 
In Section \ref{sec:hsf}, we review the nonlinear sigma model on the Grassmann manifold obtained in the harmonic superspace formalism and show that the on-shell component Lagrangian is equivalent to the on-shell component Lagrangian that is obtained in the ${\mathcal{N}}=1$ superspace formalism. 
In Section \ref{sec:psf}, we compare the potential of the complex projective space that is obtained in the projective superspace formalism with the potential that is obtained in the ${\mathcal{N}}=1$ superspace formalism and show that they are equivalent.

\subsection{${\mathcal{N}}=1$ superspace formalism}
\label{sec:n1}
\setcounter{equation}{0}

The ${\mathcal{N}}=2$ nonlinear sigma model on the Grassmann manifold is studied in the $\mathcal{N}=1$ superspace formalism in \cite{Arai:2002xa,Lindstrom:1983rt,Rocek:1980kc}. The mass-deformed nonlinear sigma model is
\begin{align}\label{ea:n=2nlsmordb}
S=&\int d^4x \Big\{\int d^4\th \Big[\tr\Big(\Ph\bar{\Ph} e^V+\bar{\Psi}\Psi e^{-V}-c V \Big)\Big] \nn \\
&+\int d^2\th \Big[\tr\Big(\Xi(\Ph\Psi-b I_M)+\Ph M\Psi\Big)+\mathrm{(c.t.)}\Big]\Big\} ,\nn\\
&(c\in \mathbf{R}_{\geq0},~ b\in \mathbf{C}).
\end{align}
We follow the convention of \cite{Galperin:2001uw}. We study the action (\ref{ea:n=2nlsmordb}) in terms of component fields. The superfields are expanded as follows:
\begin{align}\label{eq:n=1_formalism}
&\Ph_{\sss{A}}^{~\sss{I}}(y)={\mca}_{\sss{A}}^{~\sss{I}}(y)+\sqrt{2}\th\z_{\sss{A}}^{~\sss{I}}(y)+\th\th F_{\sss{A}}^{~\sss{I}}(y), \quad (y^\m=x^\m+i\th\s^\m\bar{\th}),  \nn\\
&\Psi_{\sss{I}}^{~\sss{A}}(y)={\mcb}_{\sss{I}}^{~\sss{A}}(y)+\sqrt{2}\th\e_{\sss{I}}^{~\sss{A}}(y)+\th\th G_{\sss{I}}^{~\sss{A}}(y) , \nn\\
&V_{\sss{A}}^{~\sss{B}}(x)=2\th\s^\m\bar{\th}A_{\m{\sss{A}}}^{~~\sss{B}}(x)+ i\th\th\bar{\th}\bar{\l}_{\sss{A}}^{~\sss{B}}(x)-i\bar{\th}\bar{\th}\th\l_{\sss{A}}^{~\sss{B}}(x)+\th\th\bar{\th}\bar{\th}D_{\sss{A}}^{~\sss{B}}(x), \nn\\
&\Xi_{\sss{A}}^{~\sss{B}}(y)=-{\mcs}_{\sss{A}}^{~\sss{B}}(y)+\th\o_{\sss{A}}^{~\sss{B}}(y)+\th\th K_{\sss{A}}^{~\sss{B}}(y), \nn\\
& (i=1,\cdots,N+M;~ A=1,\cdots,M).
\end{align}
The auxiliary fields are solved as
\begin{align}\label{eq:auxfld1}
&F=-\bar{\mcb} \bar{M}+ \bar{\mcs} \bar{\mcb},~~\mbox{(c.t.)}, \nn\\
&G=-\bar{M} \bar{\mca} + \bar{\mca} \bar{\mcs},~~\mbox{(c.t.)}.
\end{align}
Then the action (\ref{ea:n=2nlsmordb}) becomes
\begin{align} \label{eq:n=1_component_act}
S=\int & d^4x \tr\Big[\overline{D_\m\mca} D^\m\mca
+D_\m\mcb \overline{D^\m\mcb}
-i\bar{\z}\bar{\s}^\m D_\m\z 
-i\e \s^\m \overline{D_\m \e} \nn\\
&+\frac{\sqrt{2}i}{2}\lt(\bar{\mca}\l\z-\bar{\z}\bar{\l}\mca\rt) 
+\frac{\sqrt{2}i}{2}\lt( \mcb\bar{\l}\bar{\e} - \e\l\bar{\mcb}\rt)\nn\\
&-\frac{\sqrt{2}}{2}\lt(\e\o\mca + \mcb\o\z\rt)
-\frac{\sqrt{2}}{2}\lt(\bar{\mca}\bar{\o}\bar{\e}+\bar{\z}\bar{\o}\bar{\mcb}\rt) \nn\\
&-|\mca M -\mcs\mca|^2 - |M\mcb-\mcb \mcs|^2 
-\e\lt(\z M -\mcs\z\rt)-\bar{\z}\lt(\bar{\e}\bar{M}-\bar{\mcs}\bar{\e}\rt) \nn\\
&+(\mca\bar{\mca}- \bar{\mcb}\mcb-cI_M)D+K(\mca\mcb-bI_M)+\bar{K}(\bar{\mcb}\bar{\mca}-b^\ast I_M)
\Big],
\end{align}
with the covariant derivatives defined by 
\begin{align}\label{eq:n=1_covariant_der}
&D_\m\mca=\p_\m\mca-iA_\m \mca, \quad D_\m\mcb=\p_\m\mcb+i \mcb A_\m , \nn\\
&D_\m\z=\p_\m\z-iA_\m \z, \quad ~~ \,  \overline{D_\m\e}=\p_\m \bar{\e} - i  A_\m \bar{\e}.
\end{align}
It is shown in Section \ref{sec:hsf} that the on-shell component action (\ref{eq:n=1_component_act}) is equivalent to the on-shell component action, which is obtained from the action that is written in the harmonic superspace formalism \cite{Arai:2002xa,Galperin:2001uw,Galperin:1985dw}.

We consider the massless case of the action (\ref{ea:n=2nlsmordb}) by setting $M=0$:
\begin{align}\label{eq:ordnryfld_act}
S=\int d^4x \Big\{&\int d^4\th \Big[\tr(\Ph\bar{\Ph} e^V)+\tr(\bar{\Psi}\Psi e^{-V})-c\tr V \Big] \nn \\
&+\int d^2\th \Big[\tr\Big(\Xi(\Ph\Psi-b I_M)\Big)+\mathrm{(conjugate~transpose)}\Big]\Big\} ,\nn\\
&~(c\in \mathbf{R}_{\geq0},~ b\in \mathbf{C}). 
\end{align}
The constants $b$, $b^\ast$ and $c$ are the FI parameters.  The action is constrained by 
\begin{align}
&\Ph\bar{\Ph} e^V-e^{-V}\bar{\Psi}\Psi -cI_M=0, \label{eq:cons1_ordnryfld} \\
&\Ph\Psi-bI_M=0,\quad \mbox{(c.t.)}=0.  \label{eq:cons2_ordnryfld}
\end{align}
The potential of the model is calculated by solving $V$ from (\ref{eq:cons1_ordnryfld}) and substituting it back to the Lagrangian of (\ref{eq:ordnryfld_act}) \cite{Arai:2002xa,Lindstrom:1983rt}:
\begin{align} \label{eq:ordfldhkpot}
K=&\tr \sqrt{c^2I_M+4\Ph\bar{\Ph}\bar{\Psi}\Psi}-c\tr\ln\Big(cI_M+\sqrt{c^2I_M+ 4\Ph\bar{\Ph}\bar{\Psi}\Psi}\Big) \nn\\
&+c\tr\ln(\Ph\bar{\Ph}).
\end{align}

The constraint (\ref{eq:cons2_ordnryfld}) can be solved for $b=0$ case or $b\neq0$ case with proper gauge fixing. The two cases can be transformed to each other by $SU(2)_R$ transformations, which do not preserve the holomorphy \cite{Arai:2002xa,Lindstrom:1983rt}. We consider the parametrisation, which is discussed in \cite{Arai:2002xa}: \\

\bul $b=0$
\begin{align} \label{eq:b=0}
&\Ph=\lt(I_M~~f \rt),~~
\Psi=\lt(\begin{array}{c}
-f g \\
g
\end{array}\rt), \nn\\
& \quad (~f : M\times N,~~g: N\times M~).
\end{align}

\bul $b\neq 0$
\begin{align}  \label{eq:bneq0}
&\Ph=Q\lt(I_M ~~ s\rt),~
\Psi=\lt(\begin{array}{c}
I_M \\ t
\end{array}
\rt)Q,~~Q=\sqrt{b}\lt(I_M+ s t\rt)^{-\frac{1}{2}},\nn\\
&\quad (~s : M\times N,~~t: N\times M~).
\end{align}

The potential (\ref{eq:ordfldhkpot}) with the parametrisation (\ref{eq:b=0}) is obtained in \cite{Arai:2002xa}:
\begin{align}\label{eq:compfldkhpot}
K=&\tr\sqrt{c^2I_M+4(I_M+f\bar{f})\bar{g}(I_N+\bar{f} f)g} \nn\\
&-c\tr\ln\lt(cI_M+\sqrt{c^2I_M+4(I_M+f\bar{f})\bar{g}(I_N+\bar{f} f)g}\rt) \nn\\
&+c\tr\ln(I_M+f\bar{f}).
\end{align}
It is shown in Section \ref{sec:psf} that for $M=1$, which corresponds to the complex projective space, and $c=1$ the potential (\ref{eq:compfldkhpot}) is equivalent to the potential that is obtained in the projective superspace formalism \cite{Arai:2006gg}.

\subsection{Harmonic superspace formalism}
\label{sec:hsf}
The nonlinear sigma model on the Grassmann manifold is constructed in \cite{Galperin:2001uw,Galperin:1985dw}. The component Lagrangian is also studied in \cite{Arai:2002xa}. We review the nonlinear sigma model and present the component action in this section. We follow the convention of \cite{Galperin:2001uw}. The action is defined in the analytic superspace:
\begin{align} \label{eq:hssfldact}
&{\mathcal{S}}=-\int
d\zeta_A^{(-4)}du\mathrm{Tr}\Big[\widetilde{\ph^+}(D^{++}_A+iV^{++})\ph^++\xi^{++}V^{++}\Big], \\
&d\zeta_A^{(-4)}=d^4x_{\sss{A}} d^2\th^+ d^2\bar{\th}^+. \nn
\end{align}
We omit the subscript $A$ of $D^{++}_A$ and $x_{\sss{A}}$, which stands for the analytic basis, through this paper. Field $\ph^{+}$ is an $M \times (N+M)$ matrix and field $V^{++}$ is an $M \times M$ matrix. The coefficient $\xi^{++}=\xi^{(ij)}u_{(i}^+u_{j)}^+$ is the set of FI parameters.

The harmonic variables $u_i^+$, $u_i^-$ are defined \cite{Galperin:2001uw} by
\begin{align}
u^{+i}u_i^-=1,
\end{align}
and the integrals of products are 
\begin{align}
&\int du u_i^+u_j^-=\frac{1}{2}\ep_{ij}, \nn\\
&\int du u_i^+u_j^+u_k^-u_l^-=\frac{1}{6}(\ep_{ik}\ep_{jl}+\ep_{il}\ep_{jk}).
\end{align}
The conjugation $~\widetilde{~}~$ is the product of the ordinary complex conjugation and the antipodal map. The conjugation acts on the coefficients in the harmonic expansions, on the superspace coordinates in the central basis and on the harmonics as follows:
\begin{align}
&\widetilde{f^{i_1 \cdots i_n}}=\overline{f^{i_1 \cdots i_n}} \equiv \bar{f}_{i_1 \cdots i_n}, \nn\\
&\widetilde{\th_{\a i}} = \overline{\th_{\a i}} = \bar{\th}^i_{\dot{\a}}, \nn\\
&\widetilde{u_i^{\pm}} = u^{\pm i} \equiv \ep^{ij}u_j^{\pm}.
\end{align}

The fields and the operator expand as follows:
\begin{align}\label{eq:ham_fld_op}
&\ph^+(\z,u)=F^+(x,u)+\sqrt{2}\th^{+\a}\psi_\a(x,u)+\sqrt{2}\bar{\theta}^+_{\dot{\a}}\bar{\varphi}(x,u)^{\dot{\a}} \nn\\
&\quad\quad\quad\quad+(\th^+)^2M^-(x,u)+(\bar{\th}^+)^2N^-(x,u)+i\th^+\s^\m\bar{\th}^+A_\m^- (x,u)  \nn\\
&\quad\quad\quad\quad  
+\sqrt{2}(\th^+)^2\bar{\th}^+_{\dot{\a}}\bar{\chi}^{(-2)\dot{\a}}(x,u)+\sqrt{2}(\bar{\th}^+)^2\th^{+\a}\ta^{(-2)}_{\a}(x,u)\nn\\
&\quad\quad\quad\quad +(\th^+)^2(\bar{\th}^+)^2 D^{(-3)}(x,u),  \nn\\
&V^{++}(\z,u)=-i(\th^+)^2\S(x)+i(\bar{\th}^+)^2\bar{\S}(x)+2i\th^+\s^\m \bar{\th}^+V_\m(x) \nn\\
&\quad\quad\quad\quad ~ ~ +\sqrt{2}(\bar{\th}^+)^2\th^{+\a}\xi^{i}_{{\sss{V}}\a}(x)u_i^{-}-\sqrt{2}(\th^+)^2\bar{\th}^+_{\dot{\a}}\bar{\xi}^{\dot{\a}i}_{\sss{V}}(x)u_i^{-} \nn\\
&\quad\quad\quad\quad+3\th^{+2}\bar{\th}^{+2}D_{\sss{V}}^{ij}(x)u_i^- u_j^-, \nn \\
&D^{++}=\p^{++}-2i\th^+\s^\m\bar{\th}^+\p_\m+i(\th^+)^2(\p_5-i\p_6)-i({\bar{\th}}^+)^2(\p_5+i\p_6)\nn\\
&\quad\quad~ +\th^{+\a}\frac{\p}{\p\th^{-\a}}+\bar{\th}^{+\dot{\a}}\frac{\p}{\p\bar{\th}^{-\dot{\a}}}.
\end{align}
The vector field $V^{++}$ in (\ref{eq:ham_fld_op}) is in the Wess-Zumino gauge.
The derivatives defined in the extra dimensions are replaced by
\begin{align} \label{eq:mass_elements}
&\p_5-i\p_6=im, \nn\\
&\p_5+i\p_6=im^\ast,
\end{align}
where $m$ and $m^\ast$ are complex mass parameters for each field.

The equation of motion of (\ref{eq:hssfldact})
\bea
D^{++}\ph^+ + i V^{++}\ph^+ =0,
\eea
produces the following equations:
\begin{align} 
&\p^{++}F^+(x,u)=0, \nn \\
&\p^{++}\psi(x,u)=0, \nn \\ 
&\p^{++}\bar{\varphi}(x,u)=0, \nn\\
&\p^{++}M^-(x,u)-\lt( F^+(x,u) \mcm - \S(x,u) F^+(x,u)\rt)=0, \nn \\
&\p^{++}N^-(x,u) + \lt(F^+(x,u) \bar{\mcm} - \bar{\S}(x,u) F^+(x,u) \rt)=0, \nn \\
&\p^{++}A^-_{\m}(x,u) -2\Big(\p_\m-iV_\m(x)\Big) F^+(x,u)=0, \label{eq:hareom1}
\end{align}
and the following equations:
\begin{align}
&\p^{++}\ta^{(-2)}(x,u)+i\s^\m \Big(\p_\m -iV_\m(x)\Big)\bar{\varphi}(x,u)+\psi(x,u)\bar{\mcm}-\bar{\S}(x,u)\psi(x,u) \nn\\
&\quad+i\xi^-_{\sss{V}}(x,u)F^+(x,u)=0, \nn\\
&\p^{++}\bar{\chi}^{(-2)}(x,u)-i\bar{\s}^\m \Big(\p_\m -iV_\m (x)\Big)\psi(x,u)-\bar{\varphi}(x,u)\mcm +\S(x,u)\bar{\varphi}(x,u) \nn\\
&\quad -i\bar{\xi}_{\sss{V}}^-(x,u)F^+(x,u)=0,\nn \\
&\p^{++}D^{(-3)}+\Big(\p^\m -iV^\m(x) \Big) A_\m^{-}(x,u) -N^-(x,u)\mcm + \S(x,u) N^-(x,u) \nn\\
&\quad + M^-(x,u)\bar{\mcm} -\bar{\S}(x,u) M^-(x,u) -i\xi_{\sss{V}}^-(x,u)\psi(x,u)+i\bar{\xi}_{\sss{V}}^-(x,u)\bar{\varphi}(x,u) \nn\\
&\quad +3iD_{\sss{V}}^{(-2)}(x,u)F^+(x,u)=0. \label{eq:hareom2}
\end{align}
The matrices $\mcm$ and $\bar{\mcm}$ are the mass matrix and the conjugate transpose of it of which the elements are (\ref{eq:mass_elements}).

The equations in (\ref{eq:hareom1}) eliminate the auxiliary fields in the harmonic expansions. The equations are solved by
\begin{align}
&F^+(x,u)=f^i(x)u^+_i,\nn\\
&\psi(x,u)=\psi(x),\quad \bar{\varphi}(x,u)=\bar{\varphi}(x),  \nn\\
&M^-(x,u)= \Big( f^i(x)\mcm - \S(x) f^i(x) \Big) u_i^-, \nn\\
&N^-(x,u)= -\Big( f^i(x)\bar{\mcm} - \bar{\S}(x) f^i(x) \Big) u_i^-, \nn\\
&A_\m^-(x,u) = 2 \Big(\p_\m - iV_\m (x) \Big) f^i(x) u_i^-. \label{eq:harsol_kin}
\end{align} The equations in (\ref{eq:hareom2}) not only eliminate the auxiliary fields in the harmonic expansions but also put the physical fields on shell.

The on-shell component field action of the ${\mathcal{N}}=2$ nonlinear sigma models of the Grassmann manifold is derived by substituting the physical fields in (\ref{eq:harsol_kin}) to the action (\ref{eq:hssfldact}):
\begin{align}\label{ea:harmonic_gr}
{\mathcal{S}}=\int d^4x &\tr\Big[\overline{D^\m f_i} D_\m f^i -i\bar{\psi}\bar{\s}^\m D_\m\psi
-i\varphi \s^\m \overline{D_\m {\varphi}} \nn\\
&+\frac{i}{2}\bar{f}_i\xi_{\sss{V}}^i\psi -\frac{i}{2}\bar{f}_i\bar{\xi}_{\sss{V}}^i\bar{\varphi}
-\frac{i}{2}\bar{\psi}\bar{\xi}_{\sss{V}i}f^i -\frac{i}{2}\varphi\xi_{\sss{V}i}f^i  \nn\\
&-\frac{1}{2}\lt(\mcm\bar{f}_i-\bar{f}_i\S\rt)\lt(f^i\bar{\mcm}-\bar{\S}f^i\rt)
-\frac{1}{2}\lt(\bar{\mcm}\bar{f}_i-\bar{f}_i\bar{\S}\rt)
\lt(f^i\mcm-\S f^i\rt) \nn\\
&-\bar{\psi}\lt(\bar{\varphi}\mcm-\S\bar{\varphi}\rt)-\varphi\lt(\psi\bar{\mcm}-\bar{\S}\psi\rt) \nn\\
&+if^{(i}\bar{f}^{j)}D_{V(ij)}- \xi^{(ij)}D_{\sss{V}(ij)}\Big],
\end{align}
with the covariant derivatives defined by 
\begin{align}
&D_\m f^i=\p_\m f^i  -iV_\m f^i, \nn\\
&D_\m\psi=\p_\m \psi  -iV_\m \psi, \quad
\overline{D_\m\varphi}= \p_\m \bar{\varphi}  -iV_\m  \bar{\varphi}.
\end{align}

The action (\ref{ea:harmonic_gr}) and the action (\ref{eq:n=1_component_act}) are equivalent. They are related by the following equations:   
\begin{align} \label{eq:comp_cmpr}
&f^1=\mca,\quad f^2=\bar{\mcb},\quad \bar{f}_1=\bar{\mca},\quad \bar{f}_2=\mcb, \nn \\
&\psi=\z,\quad \bar{\varphi}=-\bar{\e}, \nn \\
&\S:=\frac{e^{i\a}}{\sqrt{2}}\lt(\S\bar{\S}+\bar{\S} \S \rt)^{\frac{1}{2}}, \quad \S=-\bar{\mcs}, \nn\\
&\bar{\S}:=\frac{e^{-i\a}}{\sqrt{2}}\lt(\S \bar{\S}+\bar{\S}\S\rt)^{\frac{1}{2}}, \quad \bar{\S}= -\mcs, \nn \\
&\mcm:=\frac{e^{i\b}}{\sqrt{2}}\lt(\mcm\bar{\mcm}+\bar{\mcm}\mcm\rt)^{\frac{1}{2}}, \quad \mcm=-\bar{M},\nn\\
&\bar{\mcm}:=\frac{e^{-i\b}}{\sqrt{2}}\lt(\mcm\bar{\mcm}+\bar{\mcm}\mcm\rt)^{\frac{1}{2}}, \quad \bar{\mcm}=-M, \nn\\
&\xi_{\sss{V}}^1=\sqrt{2}\l, \quad \xi_{\sss{V}}^2=\sqrt{2}i\o, \nn \\
&D_{\sss{V}11}=iK,\quad D_{\sss{V}22}=-i\bar{K},\quad D_{\sss{V}(12)}=-2iD, \nn \\
&\xi^{11}=-ib,\quad \xi^{22}=ib^\ast,\quad \xi^{(12)}=\frac{ic}{2}. 
\end{align}

\subsection{Projective superspace formalism}
\label{sec:psf}
The K\"{a}hler potential of the ${\mathcal{N}}=2$ nonlinear sigma model on the Grassmann manifold $G_{N+M,M}$, which is obtained in the projective superspace \cite{Arai:2006gg}, is
\begin{align}\label{eq:pothkproj}
K&=\tr\ln\lt(I_M+\bar{u} u\rt)-\tr\ln\lt(I_M+\sqrt{I_M+\tilde{v}\bar{\tilde{v}}}\rt)
+\tr\sqrt{I_M+\tilde{v}\bar{\tilde{v}}} \nn\\
&=\tr\ln\lt(I_N+u \bar{u} \rt)-\tr\ln\lt(I_N+\sqrt{I_N+\bar{\tilde{v}}\tilde{v}}\rt)
+\tr\sqrt{I_N+\bar{\tilde{v}}\tilde{v}},
\end{align}
with
\begin{align} \label{eq:proj_flds}
&\tilde{v}=s^{-1}v\underline{s}^{-1}, \nn\\
&s^{-2}=I_M+\bar{u} u, \nn\\
&\underline{s}^{-2}=I_N+u\bar{u}.
\end{align}
Field $u$ is an $N\times M$ matrix and field $v$ is an $M\times N$ matrix. 

The K\"{a}hler potential (\ref{eq:pothkproj}) and the K\"{a}hler potential (\ref{eq:compfldkhpot}) are equivalent for $M=1$, which corresponds to the complex projective space, and $c=1$. They are related by the following equations:
\begin{align}
&u = \bar{f},\quad (\mbox{c.t.}), \nn\\
&v = 2 \bar{g}, \quad (\mbox{c.t.}).
\end{align}

~~\\~~\\
\noindent {\bf Acknowledgement}

T. K. was supported by the National Research Foundation of Korea (NRF) grant funded by the Korea government (MEST) (NRF-2018R1D1A1B07051127). S. S. was supported by Basic Science Research Program through the National Research Foundation of Korea (NRF) funded by the Ministry of Education (No. NRF-2017R1D1A1B03034222).

%
%
\appendix
\def\thesection{Appendix \Alph{section}}
\setcounter{equation}{0}
\renewcommand{\theequation}{\Alph{section}.\arabic{equation}}

\section{$SO(2N)/U(N)$ with $N=4m$}  \label{app:so8m}
\setcounter{equation}{0}
The relevant elements of $\lra{A \leftrightarrow B}$, $\lra{B \leftrightarrow C}$, $\lra{C \leftrightarrow A}$ of the mass-deformed nonlinear sigma model on $SO(2N)/U(N)$ with $N=4m$ are presented in this section.

\begin{align}
&\mca_{\lra{A \leftrightarrow B}}^{({\sss{N-1}},{\sss{N-1}})}
=\frac{1}{\sqrt{s_{\sss{N-1}}}}\exp\lt(m_{\sss{N-1}}x^1+n_{\sss{N-1}}x^2+a_{{\sss{N-1};\sss{N-1}}}+ib_{{\sss{N-1};\sss{N-1}}}\rt), \nn\\
&\mca_{\lra{A \leftrightarrow B}}^{({\sss{N-1}},{\sss{2N}})}=\frac{1}{\sqrt{s_{\sss{N-1}}}}\exp\lt(-m_{\sss{N}}x^1-n_{\sss{N}}x^2+a_{{\sss{N-1};\sss{2N}}}+ib_{{\sss{N-1};\sss{2N}}}\rt), \nn\\
&s_{\sss{N-1}}=\frac{1}{c}\Big[\exp\lt(2m_{\sss{N-1}}x^1+2n_{\sss{N-1}}x^2 +2a_{\sss{N-1;N-1}}\rt)\nn\\
&\qquad\qquad\quad +\exp\lt(-2m_{\sss{N}}x^1-2n_{\sss{N}}x^2+2a_{\sss{N-1;2N}}\rt)\Big].
\end{align}

\begin{align}
&\mca_{\lra{A \leftrightarrow B}}^{({\sss{N}},{\sss{N}})}=\frac{1}{\sqrt{s_{\sss{N}}}}\exp\lt(m_{\sss{N}}x^1+n_{\sss{N}}x^2+a_{{\sss{N-1};\sss{N-1}}}+ib_{{\sss{N-1};\sss{N-1}}}\rt),\nn\\
&\mca_{\lra{A \leftrightarrow B}}^{({\sss{N}},{\sss{2N-1}})}=\frac{1}{\sqrt{s_{\sss{N}}}}\exp\lt(-m_{\sss{N-1}}x^1-n_{\sss{N-1}}x^2+a_{{\sss{N-1};\sss{2N}}}+ib_{{\sss{N-1};\sss{2N}}}+i\pi\rt), \nn\\
&s_{\sss{N}}=\frac{1}{c}\Big[\exp\lt(2m_{\sss{N}}x^1+2n_{\sss{N}}x^2 +2a_{\sss{N-1;N-1}}\rt)\nn\\
&\qquad\qquad\quad +\exp\lt(-2m_{\sss{N-1}}x^1-2n_{\sss{N-1}}x^2+2a_{\sss{N-1;2N}}\rt)\Big].
\end{align}

\begin{align}
&\mca_{\lra{B \leftrightarrow C}}^{({\sss{N-2}},{\sss{N-2}})}=\frac{1}{\sqrt{t_{\sss{N-2}}}}
\exp\lt(m_{\sss{N-2}}x^1+n_{\sss{N-2}}x^2+a_{{\sss{N-2};\sss{N-2}}}+ib_{{\sss{N-2};\sss{N-2}}}\rt), \nn\\
&\mca_{\lra{B \leftrightarrow C}}^{({\sss{N-2}},{\sss{N-1}})}=\frac{1}{\sqrt{t_{\sss{N-2}}}}
\exp\lt(m_{\sss{N-1}}x^1+n_{\sss{N-1}}x^2+a_{{\sss{N-2};\sss{N-1}}}+ib_{{\sss{N-2};\sss{N-1}}}\rt), \nn\\
&t_{\sss{N-2}}=\frac{1}{c}\Big[\exp\lt(2m_{\sss{N-2}}x^1+2n_{\sss{N-2}}x^2 +2a_{\sss{N-2;N-2}}\rt)\nn\\
&\qquad\qquad\quad+\exp\lt(2m_{\sss{N-1}}x^1+2n_{\sss{N-1}}x^2+2a_{\sss{N-2;N-1}}\rt)\Big].
\end{align}

\begin{align}
&\mca_{\lra{B \leftrightarrow C}}^{({\sss{N-1}},{\sss{2N-2}})}
=\frac{1}{\sqrt{t_{\sss{N-1}}}}\exp\lt(-m_{\sss{N-2}}x^1-n_{\sss{N-2}}x^2+a_{{\sss{N-2};\sss{N-1}}}+ib_{{\sss{N-2};\sss{N-1}}}+i\pi\rt), \nn\\
&\mca_{\lra{B \leftrightarrow C}}^{({\sss{N-1}},{\sss{2N-1}})}
=\frac{1}{\sqrt{t_{\sss{N-1}}}}\exp\lt(-m_{\sss{N-1}}x^1-n_{\sss{N-1}}x^2+a_{{\sss{N-2};\sss{N-2}}}+ib_{{\sss{N-2};\sss{N-2}}}\rt), \nn\\
&t_{\sss{N-1}}=\frac{1}{c}\Big[\exp\lt(-2m_{\sss{N-2}}x^1-2n_{\sss{N-2}}x^2+2a_{\sss{N-2;N-1}}\rt)\nn\\
&\qquad\qquad\quad  +\exp\lt(-2m_{\sss{N-1}}x^1-2n_{\sss{N-1}}x^2+2a_{\sss{N-2;N-2}}\rt)\Big].
\end{align}

\begin{align}
&\mca_{\lra{C \leftrightarrow A}}^{({\sss{N-2}},{\sss{N-2}})}
=\frac{1}{\sqrt{u_{\sss{N-2}}}}\exp\lt(m_{\sss{N-2}}x^1+n_{\sss{N-2}}x^2+a_{{\sss{N-2};\sss{N-2}}}+ib_{{\sss{N-2};\sss{N-2}}}\rt), \nn\\
&\mca_{\lra{C \leftrightarrow A}}^{({\sss{N-2}},{\sss{2N}})}
=\frac{1}{\sqrt{u_{\sss{N-2}}}}\exp\lt(-m_{\sss{N}}x^1-n_{\sss{N}}x^2+a_{{\sss{N-2};\sss{2N}}}+ib_{{\sss{N-2};\sss{2N}}}\rt), \nn\\
&u_{\sss{N-2}}=\frac{1}{c}\Big[\exp\lt(2m_{\sss{N-2}}x^1+2n_{\sss{N-2}}x^2 +2a_{\sss{N-2;N-2}}\rt) \nn\\
&\qquad\qquad\quad +\exp\lt(-2m_{\sss{N}}x^1-2n_{\sss{N}}x^2+2a_{\sss{N-2;2N}}\rt)\Big].
\end{align}

\begin{align}
&\mca_{\lra{C \leftrightarrow A}}^{({\sss{N}},{\sss{N}})}
=\frac{1}{\sqrt{u_{\sss{N}}}}\exp\lt(m_{\sss{N}}x^1+n_{\sss{N}}x^2+a_{{\sss{N-2};\sss{N-2}}}+ib_{{\sss{N-2};\sss{N-2}}}\rt), \nn\\
&\mca_{\lra{C \leftrightarrow A}}^{({\sss{N}},{\sss{2N-2}})}
=\frac{1}{\sqrt{u_{\sss{N}}}}\exp\lt(-m_{\sss{N-2}}x^1-n_{\sss{N-2}}x^2+a_{{\sss{N-2};\sss{2N}}}+ib_{{\sss{N-2};\sss{2N}}}+i\pi\rt), \nn\\
&u_{\sss{N}}=\frac{1}{c}\Big[\exp\lt(2m_{\sss{N}}x^1+2n_{\sss{N}}x^2 +2a_{\sss{N-2;N-2}}\rt)\nn\\
&\qquad\qquad\quad+\exp\lt(-2m_{\sss{N-2}}x^1-2n_{\sss{N-2}}x^2+2a_{\sss{N-2;2N}}\rt)\Big].
\end{align}

\section{$SO(2N)/U(N)$ with $N=4m+1$}  \label{app:so8mp2}
\setcounter{equation}{0}
The relevant elements of $\lra{H \leftrightarrow L}$, $\lra{L \leftrightarrow P}$, $\lra{P \leftrightarrow H}$ of the mass-deformed nonlinear sigma model on $SO(2N)/U(N)$ with $N=4m+1$ are presented in this section.

\begin{align}
&\mca_{\lra{H \leftrightarrow L}}^{({\sss{N-4}},{\sss{N-4}})} =
\frac{1}{\sqrt{p_{\sss{N-4}}}}\exp\lt(m_{\sss{N-4}}x^1+n_{\sss{N-4}}x^2+a_{{\sss{N-4};\sss{N-4}}}+ib_{{\sss{N-4};\sss{N-4}}}\rt), \nn\\
&\mca_{\lra{H \leftrightarrow L}}^{({\sss{N-4}},{\sss{N-3}})}
=\frac{1}{\sqrt{p_{\sss{N-4}}}}\exp\lt(m_{\sss{N-3}}x^1+n_{\sss{N-3}}x^2+a_{{\sss{N-4};\sss{N-3}}}+ib_{{\sss{N-4};\sss{N-3}}}\rt), \nn\\
&p_{\sss{N-4}}=\frac{1}{c}\Big[\exp\lt(2m_{\sss{N-4}}x^1+2n_{\sss{N-4}}x^2 +2a_{\sss{N-4;N-4}}\rt) \nn\\
&\qquad\qquad\quad+\exp\lt(2m_{\sss{N-3}}x^1+2n_{\sss{N-3}}x^2+2a_{\sss{N-4;N-3}}\rt)\Big].
\end{align}

\begin{align}
&\mca_{\lra{H \leftrightarrow L}}^{({\sss{N-3}},{\sss{2N-4}})}
=\frac{1}{\sqrt{p_{\sss{N-3}}}}\exp\lt(-m_{\sss{N-4}}x^1-n_{\sss{N-4}}x^2+a_{{\sss{N-4};\sss{N-3}}}+ib_{{\sss{N-4};\sss{N-3}}}+i\pi\rt), \nn\\
&\mca_{\lra{H \leftrightarrow L}}^{({\sss{N-3}},{\sss{2N-3}})}
=\frac{1}{\sqrt{p_{\sss{N-3}}}}\exp\lt(-m_{\sss{N-3}}x^1-n_{\sss{N-3}}x^2+a_{{\sss{N-4};\sss{N-4}}}+ib_{{\sss{N-4};\sss{N-4}}}\rt), \nn\\
&p_{\sss{N-3}}=\frac{1}{c}\Big[\exp\lt(-2m_{\sss{N-4}}x^1-2n_{\sss{N-4}}x^2 +2a_{\sss{N-4;N-3}}\rt)\nn\\
&\qquad\qquad\quad+\exp\lt(-2m_{\sss{N-3}}x^1-2n_{\sss{N-3}}x^2+2a_{\sss{N-4;N-4}}\rt)\Big].
\end{align}

\begin{align}
&\mca_{\lra{L \leftrightarrow P}}^{({\sss{N-3}},{\sss{N-3}})}
=\frac{1}{\sqrt{q_{\sss{N-3}}}}\exp\lt(m_{\sss{N-3}}x^1+n_{\sss{N-3}}x^2+a_{{\sss{N-3};\sss{N-3}}}+ib_{{\sss{N-3};\sss{N-3}}}\rt), \nn\\
&\mca_{\lra{L \leftrightarrow P}}^{({\sss{N-3}},{\sss{N-2}})}
=\frac{1}{\sqrt{q_{\sss{N-3}}}}\exp\lt(m_{\sss{N-2}}x^1+n_{\sss{N-2}}x^2+a_{{\sss{N-3};\sss{N-2}}}+ib_{{\sss{N-3};\sss{N-2}}}\rt), \nn\\
&q_{\sss{N-3}}=\frac{1}{c}\Big[\exp\lt(2m_{\sss{N-3}}x^1+2n_{\sss{N-3}}x^2 +2a_{\sss{N-3;N-3}}\rt)\nn\\
&\qquad\qquad\quad+\exp\lt(2m_{\sss{N-2}}x^1+2n_{\sss{N-2}}x^2+2a_{\sss{N-3;N-2}}\rt)\Big].
\end{align}

\begin{align}
&\mca_{\lra{L \leftrightarrow P}}^{({\sss{N-2}},{\sss{2N-3}})}
=\frac{1}{\sqrt{q_{\sss{N-2}}}}\exp\lt(-m_{\sss{N-3}}x^1-n_{\sss{N-3}}x^2+a_{{\sss{N-3};\sss{N-2}}}+ib_{{\sss{N-3};\sss{N-2}}}+i\pi\rt), \nn\\
&\mca_{\lra{L \leftrightarrow P}}^{({\sss{N-2}},{\sss{2N-2}})}
=\frac{1}{\sqrt{q_{\sss{N-2}}}}\exp\lt(-m_{\sss{N-2}}x^1-n_{\sss{N-2}}x^2+a_{{\sss{N-3};\sss{N-3}}}+ib_{{\sss{N-3};\sss{N-3}}}\rt), \nn\\
&q_{\sss{N-2}}=\frac{1}{c}\Big[\exp\lt(-2m_{\sss{N-3}}x^1-2n_{\sss{N-3}}x^2 +2a_{\sss{N-3;N-2}}\rt)\nn\\
&\qquad\qquad\quad+\exp\lt(-2m_{\sss{N-2}}x^1-2n_{\sss{N-2}}x^2+2a_{\sss{N-3;N-3}}\rt)\Big].
\end{align}

\begin{align}
&\mca_{\lra{P \leftrightarrow H}}^{({\sss{N-4}},{\sss{N-4}})}
=\frac{1}{\sqrt{r_{\sss{N-4}}}}\exp\lt(m_{\sss{N-4}}x^1+n_{\sss{N-4}}x^2+a_{{\sss{N-4};\sss{N-4}}}+ib_{{\sss{N-4};\sss{N-4}}}\rt),\nn\\
&\mca_{\lra{P \leftrightarrow H}}^{({\sss{N-4}},{\sss{N-2}})}
=\frac{1}{\sqrt{r_{\sss{N-4}}}}\exp\lt(m_{\sss{N-2}}x^1+n_{\sss{N-2}}x^2+a_{{\sss{N-4};\sss{N-2}}}+ib_{{\sss{N-4};\sss{N-2}}}\rt),\nn\\
&r_{\sss{N-4}}=\frac{1}{c}\Big[\exp\lt(2m_{\sss{N-4}}x^1+2n_{\sss{N-4}}x^2 +2a_{\sss{N-4;N-4}}\rt)\nn\\
&\qquad\qquad\quad +\exp\lt(2m_{\sss{N-2}}x^1+2n_{\sss{N-2}}x^2+2a_{\sss{N-4;N-2}}\rt)\Big].
\end{align}

\begin{align}
&\mca_{\lra{P \leftrightarrow H}}^{({\sss{N-2}},{\sss{2N-4}})}
=\frac{1}{\sqrt{r_{\sss{N-2}}}}\exp\lt(-m_{\sss{N-4}}x^1-n_{\sss{N-4}}x^2+a_{{\sss{N-4};\sss{N-2}}}+ib_{{\sss{N-4};\sss{N-2}}}+i\pi\rt),\nn\\
&\mca_{\lra{P \leftrightarrow H}}^{({\sss{N-2}},{\sss{2N-2}})}
=\frac{1}{\sqrt{r_{\sss{N-2}}}}\exp\lt(-m_{\sss{N-2}}x^1-n_{\sss{N-2}}x^2+a_{{\sss{N-4};\sss{N-4}}}+ib_{{\sss{N-4};\sss{N-4}}}\rt),\nn\\
&r_{\sss{N-2}}=\frac{1}{c}\Big[\exp\lt(-2m_{\sss{N-4}}x^1-2n_{\sss{N-4}}x^2 +2a_{\sss{N-4;N-2}}\rt)\nn\\
&\qquad\qquad\quad +\exp\lt(-2m_{\sss{N-2}}x^1-2n_{\sss{N-2}}x^2+2a_{\sss{N-4;N-4}}\rt)\Big].
\end{align}

\section{$Sp(N)/U(N)$ with $N=4m-1$}  \label{app:sp4mm1}
\setcounter{equation}{0}
The relevant elements of $\lra{A \leftrightarrow B}$, $\lra{B \leftrightarrow C}$, $\lra{C \leftrightarrow A}$ of the mass-deformed nonlinear sigma model on $Sp(N)/U(N)$ with $N=4m-1$ are presented in this section.

\begin{align}
&\mca_{\lra{A \leftrightarrow B}}^{({\sss{N}},{\sss{N}})}
=\frac{1}{\sqrt{s_{\sss{N}}}}\exp\lt(m_{\sss{N}}x^1+n_{\sss{N}}x^2+a_{{\sss{N};\sss{N}}}+ib_{{\sss{N};\sss{N}}}\rt),\nn\\
&\mca_{\lra{A \leftrightarrow B}}^{({\sss{N}},{\sss{2N}})}
=\frac{1}{\sqrt{s_{\sss{N}}}}\exp\lt(-m_{\sss{N}}x^1-n_{\sss{N}}x^2+a_{{\sss{N};\sss{2N}}}+ib_{{\sss{N};\sss{2N}}}\rt), \nn\\
&s_{\sss{N}}=\frac{1}{c}\Big[\exp\lt(2m_{\sss{N}}x^1+2n_{\sss{N}}x^2 +2a_{\sss{N;N}}\rt)\nn\\
&\qquad\qquad\quad+\exp\lt(-2m_{\sss{N}}x^1-2n_{\sss{N}}x^2+2a_{\sss{N;2N}}\rt)\Big].
\end{align}

\begin{align}
&\mca_{\lra{B \leftrightarrow C}}^{({\sss{N-1}},{\sss{N-1}})} 
=\frac{1}{\sqrt{t_{\sss{N-1}}}}\exp\lt(m_{\sss{N-1}}x^1+n_{\sss{N-1}}x^2+a_{{\sss{N-1};\sss{N-1}}}+ib_{{\sss{N-1};\sss{N-1}}}\rt), \nn\\
&\mca_{\lra{B \leftrightarrow C}}^{({\sss{N-1}},{\sss{N}})} 
=\frac{1}{\sqrt{t_{\sss{N-1}}}}\exp\lt(m_{\sss{N}}x^1+n_{\sss{N}}x^2+a_{{\sss{N-1};\sss{N}}}+ib_{{\sss{N-1};\sss{N}}}\rt), \nn\\
&t_{\sss{N-1}}=\frac{1}{c}\Big[\exp\lt(2m_{\sss{N-1}}x^1+2n_{\sss{N-1}}x^2 +2a_{\sss{N-1;N-1}}\rt)\nn\\
&\qquad\qquad\quad +\exp\lt(2m_{\sss{N}}x^1+2n_{\sss{N}}x^2+2a_{\sss{N-1;N}}\rt)\Big].
\end{align}

\begin{align}
&\mca_{\lra{B \leftrightarrow C}}^{({\sss{N}},{\sss{2N-1}})}
=\frac{1}{\sqrt{t_{\sss{N}}}}\exp\lt(-m_{\sss{N-1}}x^1-n_{\sss{N-1}}x^2+a_{{\sss{N-1};\sss{N}}}+ib_{{\sss{N-1};\sss{N}}}+i\pi\rt), \nn\\
&\mca_{\lra{B \leftrightarrow C}}^{({\sss{N}},{\sss{2N}})}
=\frac{1}{\sqrt{t_{\sss{N}}}}\exp\lt(-m_{\sss{N}}x^1-n_{\sss{N}}x^2+a_{{\sss{N-1};\sss{N-1}}}+ib_{{\sss{N-1};\sss{N-1}}}\rt), \nn\\
&t_{\sss{N}}=\frac{1}{c}\Big[\exp\lt(-2m_{\sss{N-1}}x^1-2n_{\sss{N-1}}x^2 +2a_{\sss{N-1;N}}\rt)\nn\\
&\qquad\qquad\quad +\exp\lt(-2m_{\sss{N}}x^1-2n_{\sss{N}}x^2+2a_{\sss{N-1;N-1}}\rt)\Big].
\end{align}

\begin{align}
&\mca_{\lra{C \leftrightarrow A}}^{({\sss{N-1}},{\sss{N-1}})}
=\frac{1}{\sqrt{u_{\sss{N-1}}}}\exp\lt(m_{\sss{N-1}}x^1+n_{\sss{N-1}}x^2+a_{{\sss{N-1};\sss{N-1}}}+ib_{{\sss{N-1};\sss{N-1}}}\rt), \nn\\
&\mca_{\lra{C \leftrightarrow A}}^{({\sss{N-1}},{\sss{2N-1}})}
=\frac{1}{\sqrt{u_{\sss{N-1}}}}\exp\lt(-m_{\sss{N-1}}x^1-n_{\sss{N-1}}x^2+a_{{\sss{N-1};\sss{2N-1}}}+ib_{{\sss{N-1};\sss{2N-1}}}\rt), \nn\\
&u_{\sss{N-1}}=\frac{1}{c}\Big[\exp\lt(2m_{\sss{N-1}}x^1+2n_{\sss{N-1}}x^2 +2a_{\sss{N-1;N-1}}\rt)\nn\\
&\qquad\qquad\quad +\exp\lt(-2m_{\sss{N-1}}x^1-2n_{\sss{N-1}}x^2+2a_{\sss{N-1;2N-1}}\rt)\Big].
\end{align}

\section{$Sp(N)/U(N)$ with $N=4m$}  \label{app:sp4m}
\setcounter{equation}{0}
The relevant elements of $\lra{H \leftrightarrow L}$, $\lra{L \leftrightarrow P}$, $\lra{P \leftrightarrow H}$ of the mass-deformed nonlinear sigma model on $Sp(N)/U(N)$ with $N=4m$ are presented in this section.

\begin{align}
&\mca_{\lra{H \leftrightarrow L}}^{({\sss{N-3}},{\sss{N-3}})} 
=\frac{1}{\sqrt{p_{\sss{N-3}}}}\exp\lt(m_{\sss{N-3}}x^1+n_{\sss{N-3}}x^2+a_{{\sss{N-3};\sss{N-3}}}+ib_{{\sss{N-3};\sss{N-3}}}\rt), \nn\\
&\mca_{\lra{H \leftrightarrow L}}^{({\sss{N-3}},{\sss{N-2}})} 
=\frac{1}{\sqrt{p_{\sss{N-3}}}}\exp\lt(m_{\sss{N-2}}x^1+n_{\sss{N-2}}x^2+a_{{\sss{N-3};\sss{N-2}}}+ib_{{\sss{N-3};\sss{N-2}}}\rt), \nn\\
&p_{\sss{N-3}}=\frac{1}{c}\Big[\exp\lt(2m_{\sss{N-3}}x^1+2n_{\sss{N-3}}x^2 +2a_{\sss{N-3;N-3}}\rt)\nn\\
&\qquad\qquad\quad  +\exp\lt(2m_{\sss{N-2}}x^1+2n_{\sss{N-2}}x^2+2a_{\sss{N-3;N-2}}\rt)\Big].
\end{align}

\begin{align}
&\mca_{\lra{H \leftrightarrow L}}^{({\sss{N-2}},{\sss{2N-3}})}
=\frac{1}{\sqrt{p_{\sss{N-2}}}}\exp\lt(-m_{\sss{N-3}}x^1-n_{\sss{N-3}}x^2+a_{{\sss{N-3};\sss{N-2}}}+ib_{{\sss{N-3};\sss{N-2}}}+i\pi\rt), \nn\\
&\mca_{\lra{H \leftrightarrow L}}^{({\sss{N-2}},{\sss{2N-2}})}
=\frac{1}{\sqrt{p_{\sss{N-2}}}}\exp\lt(-m_{\sss{N-2}}x^1-n_{\sss{N-2}}x^2+a_{{\sss{N-3};\sss{N-3}}}+ib_{{\sss{N-3};\sss{N-3}}}\rt), \nn\\
&p_{\sss{N-2}}=\frac{1}{c}\Big[\exp\lt(-2m_{\sss{N-3}}x^1-2n_{\sss{N-3}}x^2+2a_{\sss{N-3;N-2}}\rt) \nn\\
&\qquad\qquad\quad  +\exp\lt(-2m_{\sss{N-2}}x^1-2n_{\sss{N-2}}x^2 +2a_{\sss{N-3;N-3}}\rt)\Big].
\end{align}

\begin{align}
&\mca_{\lra{L \leftrightarrow P}}^{({\sss{N-2}},{\sss{N-2}})}
=\frac{1}{\sqrt{q_{\sss{N-2}}}}\exp\lt(m_{\sss{N-2}}x^1+n_{\sss{N-2}}x^2+a_{{\sss{N-2};\sss{N-2}}}+ib_{{\sss{N-2};\sss{N-2}}}\rt), \nn\\
&\mca_{\lra{L \leftrightarrow P}}^{({\sss{N-2}},{\sss{N-1}})}
=\frac{1}{\sqrt{q_{\sss{N-2}}}}\exp\lt(m_{\sss{N-1}}x^1+n_{\sss{N-1}}x^2+a_{{\sss{N-2};\sss{N-1}}}+ib_{{\sss{N-2};\sss{N-1}}}\rt), \nn\\
&q_{\sss{N-2}}=\frac{1}{c}\Big[\exp\lt(2m_{\sss{N-2}}x^1+2n_{\sss{N-2}}x^2 +2a_{\sss{N-2;N-2}}\rt) \nn\\
&\qquad\qquad\quad  +\exp\lt(2m_{\sss{N-1}}x^1+2n_{\sss{N-1}}x^2+2a_{\sss{N-2;N-1}}\rt)\Big].
\end{align}

\begin{align}
&\mca_{\lra{L \leftrightarrow P}}^{({\sss{N-1}},{\sss{2N-2}})}
=\frac{1}{\sqrt{q_{\sss{N-1}}}}\exp\lt(-m_{\sss{N-2}}x^1-n_{\sss{N-2}}x^2+a_{{\sss{N-2};\sss{N-1}}}+ib_{{\sss{N-2};\sss{N-1}}}+i\pi\rt), \nn\\
&\mca_{\lra{L \leftrightarrow P}}^{({\sss{N-1}},{\sss{2N-1}})} 
=\frac{1}{\sqrt{q_{\sss{N-1}}}}\exp\lt(-m_{\sss{N-1}}x^1-n_{\sss{N-1}}x^2+a_{{\sss{N-2};\sss{N-2}}}+ib_{{\sss{N-2};\sss{N-2}}}\rt), \nn\\
&q_{\sss{N-1}}=\frac{1}{c}\Big[\exp\lt(-2m_{\sss{N-2}}x^1-2n_{\sss{N-2}}x^2 +2a_{\sss{N-2;N-1}}\rt) \nn\\
&\qquad\qquad\quad  +\exp\lt(-2m_{\sss{N-1}}x^1-2n_{\sss{N-1}}x^2+2a_{\sss{N-2;N-2}}\rt)\Big].
\end{align}

\begin{align}
&\mca_{\lra{P \leftrightarrow H}}^{({\sss{N-3}},{\sss{N-3}})}
=\frac{1}{\sqrt{r_{\sss{N-3}}}}\exp\lt(m_{\sss{N-3}}x^1+n_{\sss{N-3}}x^2+a_{{\sss{N-3};\sss{N-3}}}+ib_{{\sss{N-3};\sss{N-3}}}\rt),\nn \\
&\mca_{\lra{P \leftrightarrow H}}^{({\sss{N-3}},{\sss{N-1}})} 
=\frac{1}{\sqrt{r_{\sss{N-3}}}}\exp\lt(m_{\sss{N-1}}x^1+n_{\sss{N-1}}x^2+a_{{\sss{N-3};\sss{N-1}}}+ib_{{\sss{N-3};\sss{N-1}}}\rt), \nn\\
&r_{\sss{N-3}}=\frac{1}{c}\Big[\exp\lt(2m_{\sss{N-3}}x^1+2n_{\sss{N-3}}x^2 +2a_{\sss{N-3;N-3}}\rt)\nn\\
&\qquad\qquad\quad +\exp\lt(2m_{\sss{N-1}}x^1+2n_{\sss{N-1}}x^2+2a_{\sss{N-3;N-1}}\rt)\Big].
\end{align}

\begin{align}
&\mca_{\lra{P \leftrightarrow H}}^{({\sss{N-1}},{\sss{2N-3}})}
=\frac{1}{\sqrt{r_{\sss{N-1}}}}\exp\lt(-m_{\sss{N-3}}x^1-n_{\sss{N-3}}x^2+a_{{\sss{N-3};\sss{N-1}}}+ib_{{\sss{N-3};\sss{N-1}}}+i\pi\rt), \nn\\
&\mca_{\lra{P \leftrightarrow H}}^{({\sss{N-1}},{\sss{2N-1}})}
=\frac{1}{\sqrt{r_{\sss{N-1}}}}\exp\lt(-m_{\sss{N-1}}x^1-n_{\sss{N-1}}x^2+a_{{\sss{N-3};\sss{N-3}}}+ib_{{\sss{N-3};\sss{N-3}}}\rt), \nn\\
&r_{\sss{N-1}}=\frac{1}{c}\Big[\exp\lt(-2m_{\sss{N-3}}x^1-2n_{\sss{N-3}}x^2 +2a_{\sss{N-3;N-1}}\rt)\nn\\
&\qquad\qquad\quad +\exp\lt(-2m_{\sss{N-1}}x^1-2n_{\sss{N-1}}x^2+2a_{\sss{N-3;N-3}}\rt)\Big].
\end{align}

\end{document}